\documentstyle[12pt]{article}
\input psfig
\begin{document}

\smallskip

\begin{center}
{\bf  Pair correlations and magnetic susceptibility of small Al-grains\\}
\vspace{5mm}
{N.K. Kuzmenko$^{1)}$, V.M. Mikhajlov$^{2)}$\\
and  S.Frauendorf$^{3)}$}\\
{\em ${}^{1)}$V.G. Khlopin Radium Institute, 194021,
St.-Petersburg, Russia } \\
{\em ${}^{2)}$Institute of Physics St.--Petersburg State
University 198904, Russia}\\
{\em ${}^{3)}$IKHP, Forschungszentrum Rossendorf
e.V., PF 510119, 01314 Dresden, Germany }\\
\end{center}

\vspace{2cm}
{\large \bf Abstract}

Pair correlations and the magnetic susceptibility of electrons
in a spherical cavity are studied both for grand canonical and the 
canonical ensemble. The coupling constant of the $BCS$ Hamiltonian
is adjusted to experimental values of the gap parameter.
The gap parameter is found to increase for small grains as a consequence
of the pronounced  shell structure in the spectrum
of  the spherical cavity. The sharp phase transition  at $T_c$ 
is smeared out   for the canonical ensemble. 
The strong paramagnetic susceptibility of the normal electrons
in the cavity is reduced by the superconductivity, but it remains positive.

\newpage
 \section{Introduction}
   Whereas many properties of superconducting bulk
 metals at low temperatures have been well studied, the influence
 of size effects on the same properties of small metal grains or
 clusters is still a problem of a current interest. The minimal size of
 grains in which the phenomenon of superconductivity can be
   observed has not been not established by now. It has been known
  for about three decades that 
   metal grains on the  nanometer-scale  display superconductivity
\cite{1}.
 
The  focus of the paper are the electronic 
 finite size effects in the grains, which  reveal
 themselves not only in the discreteness of the energies
 of the  electron states but also in their shell structure. This means,
that the level density strongly fluctuates as function of the energy.
In regions with high level density the pair correlations are enhanced.
This  kind  of enhancement has been demonstrated
 for electrons confined to a slab \cite{9} and to a cube \cite{11}. 
In this paper we are going to investigate the same  effect
 as well as the magnetic response of small grains 
 by means of yet another very simple model.  It is 
assumed  that the electrons are confined inside  a sphere or hemisphere with a
perfect surface and that only the electrons on the highly degenerated
Fermi level take part in the pair correlations.  
The spherical model has a very pronounced shell structure. In this
respect it contrasts  the extended Landau - Ginzburg model for grains
 \cite{22}, which does not
take  into account any shell structure.
Comparing these two extremes will provide insight
into the properties of real grains with electronic shell structure.
Though the model   is a very strong idealization of realistic grains 
it has the advantage that   
the  exact solutions of the many body problem
with finite particle number are known. This permits to exactly
calculate
finite size effects due the conservation of the electron number
 as well as the enhancement of the 
magnetic response in a finite system. It also allows to judge   
the typical approximations (mean field approximation, grand canonical ensemble)
one has to resort to in more realistic models  of the electron system of
the grain. 
In particular,  it will be demonstrated that the strong shell structure of the (hemi)
sphere  generates a 
dramatic enhancement of the pair correlations with decreasing size,
resulting in a growth of the transition temperature.     

 The  experimental  transition temperatures
 $T_c(N)$ for grains with $N$ delocalized electrons turn out to be
 larger  than or equal to the bulk value.  For Al, a weak-coupling 
superconducting metal, the
measured ratio
 $f=T_c(N)/T_c(bulk)$  grows   considerably
  with decreasing $N$: $f\approx 1.5$ for $N\sim 10^5$
  and $f\approx 3$ for  $N\sim 10^3$   \cite{2,3,4}.
 Grains of intermediate-coupling metals show a more
 moderate growth of $T_c(N)$:  $f\approx 1.2$  for In grains with
 $N\sim$ $10^5$ \cite{2},  $f\approx 1.1$ for Sn grains  with
 $N\sim$ $10^6$ \cite{5}. For strong-coupling
 Pb the transition temperature $T_c(N)$ is about equal to the bulk value
 down to grains with $N\sim 10^3$ \cite{2}. 
 Recently,
  Black, Ralf and Tinkham \cite{6,7} provided additional evidence.
By tunneling experiments on grains of $R\sim$ (5-10) nm,
 the pairing gap $\Delta$ has been measured 
to be as large as $\sim 2$ times the bulk value.

The increase of
 the transition temperature and the pairing gap
  has  been  attributed to
 an increase of the coupling constant $\lambda$
with decreasing size of the grains.
 Various theoretical models have been suggested for
 explanation of such an increase \cite{2,8,10,12,13,14,15,16}.
In this work we do not investigate  the possible mechanisms
behind such a change of the coupling constant. Rather, 
we choose a phenomenological approach, introducing an explicit
dependence of the effective interaction strength $\lambda$
on the electron number $N$.
We are going to demonstrate that
in  studying the $N$ - dependence of the coupling constant the
shell structure has to be taken into account.

 The BCS  theory of our model is exposed 
in sections \ref{bcs} and \ref{average}.
 We limit ourselves to  Al grains
 with $10^3<N<10^5$, for which
 the best experimental data on $\Delta$ are available.
 In section \ref{g}., these are used to determine
 $G(N)$ 
by  comparing the  empirical and calculated values of $T_c(N)$.

 It is known \cite{5,17} that the specific heat $C(T)$ of
 superconducting metallic grains does not show the  singularity
 at $T_c$ seen  in bulk metals.  For  grains, the function $C(T)$
has a maximum at $T_0$ that is smaller than $T_c$ for bulk metals.
 With decreasing
 grain size this maximum  shifts to smaller
 temperature and the width of the peak becomes wider.
The question arises, how to define a transition temperature $T_c$
from the non-singular function $C(T)$.
Calculations in the grand canonical BCS  approximation
cannot reproduce the observed $T$-dependence of $C$.  Therefore, in section
\ref{can} the influence
 of the superconductivity on $C(T)$ is studied within the canonical
ensemble. The results suggest a definition of $T_c$, which coincides with the
one in $BCS$ theory as well as  with previous definitions, introduced
for interpreting the measurements of $C(T)$
\cite{5,17} and electromagnetic properties \cite{16,17} in grains.

Superconducting bulk metals show
the Meissner effect.
The diamagnetic susceptibility takes its
 maximum,  compensating completely  external magnetic
field inside the metal. For grains with sizes comparable or smaller than
the coherence and penetration  lengths an incomplete compensation is
expected, which will be investigated in section \ref{mag}.
The issue is complicated by the fact, that in normal (non-superconducting)
grains the susceptibility  at low temperatures strongly deviates from its
bulk values. Both para- and diamagnetic enhancements appear a certain
electron numbers, reflecting the electronic shell structure of the grain.
This has  recently been
studied in \cite{richter,cm,supershell}, where the references to further work can be found.
Hence, an intricate interplay between the paramagnetism due to the
shell structure and the diamagnetism due to the superconductivity is
expected for superconducting grains, which will  also be studied in
section \ref{mag}.

\section{The spherical cavity model in $BCS$ approximation}\label{bcs}

We consider $N$ electrons in a cavity of
radius  $R=r_0N^{1/3}$, i. e. 
we employ the spherically symmetric rectangular well with
   infinite walls  as a simplified model of the  mean
   field that confines the delocalized electrons of the grain.
  The  energies $e_t$ are given by the roots of the
 Bessel functions. 

Thus, we describe the system of $N$ electrons by
means of  the effective electron Hamiltonian
\begin{equation}\label{eq:hpair}
H-\mu N=\sum (e_t-\mu)a^+_ta_t -
G(N)\sum a^+_ta^+_{\bar {t}} a_{\bar {s}}a_s.
\end{equation}                                      
The interaction strength $G$ is treated as a parameter that is
fixed by comparison with the experimental pair gaps $\Delta(N)$
measured for different grain sizes.  
 The pairing interaction in (1) acts  only among
 electron levels inside the  interval $|e_t-\mu|<\hbar \omega_D$,
 where the Debye frequency  is taken to be
 the same as in the bulk.

The treatment of the Hamiltonian (\ref{eq:hpair}) in $BCS$ mean field
approximation is a standard problem in nuclear physics. For example it
is exposed in the textbook \cite{20}. We have solved the problem 
numerically.
If a grain contains  $10^3$ or more delocalized electrons
the single-particle spectrum posses a high degeneracy in the
vicinity of the Fermi level ($F$). The average orbital momentum
$l$ is of order $N^{1/3}$.  Each level can be occupied by
about  $4\cdot N^{1/3}$ electrons. The number of these degenerated levels
inside the Debye interval $2\hbar \omega_D$ varies from 1 ($N\sim 10^3$) to
5 ($N\sim 10^5$). However, as demonstrated in fig. 1,
practically it is sufficient to take into account
    the Fermi level alone because for 
  particle numbers ($10^3<N<10^5$) the distance to
   next levels is always much larger than the
 pairing gap $\Delta$, found by means of the full  $BCS$
 equations.
The isolation the Fermi
level and its high degeneracy are the conditions to apply the 
{\em single shell
 model} \cite{20} for pairing.  This approximation takes into account
only the pair  interaction among  the electrons in
the partially filled Fermi level.   
In the grand canonical $BCS$ approximation, the single shell model was used in
   \cite{21} for a half filled shell. We consider the case of
  arbitrary numbers of particles in the shell and, in sections \ref{can}
and \ref{mag}, the canonical ensemble of the exact many body states.

In the  single shell approximation, which we are going to study, the 
spherical cavity is equivalent with a hemisphere. This is seen as follows.
The wave functions in the spherical cavity have good orbital angular momentum
$l$. All wave functions with odd $l$ are equal to zero in the the plane $z=0$.
Thus, they fulfill also the boundary condition for 
the hemisphere and are ( with the appropriate choice of the normalization) 
the wave function of this type of cavity. None of the conclusions 
drawn below depends on $l$ being odd or even. Thus, they are valid for the 
hemisphere as well. Clusters on surfaces, the
superconductivity of which is studied experimentally, have the shape of
somewhat flattened hemispheres. 

   The single shell approximation leads to  considerable simplifications
 because there are no summations over single particle states in
 the $BCS$ equations and many
 quantities become  analytical expressions of the temperature and
the filling parameter
\begin{equation}
n=\frac{N_{sh}}{2M},~~ M=2l+1,
\end{equation}                              
 where $N_{sh}$ is  the actual
and  $2(2l+1)$  the maximal number of  electrons on the Fermi level, which
has $M=2l+1$ magnetic substates and 2 orientations of the spin.
There is only one quasiparticle energy
\begin{equation}\label{e:qp}
E=\sqrt{(e-\mu)^2+\Delta^2},
\end{equation}                      
which  is independent of $n$  at zero temperature,
\begin{equation}\label{e:e0}
E(0)=\frac{GM}{2}.
\end{equation}                                      
At finite temperature the  quasiparticle energy
is obtained as the solution of the implicit equation
\begin{equation}\label{e:gap} E=E(0)\tanh\{\frac{E}{2T}\}.
\end{equation}                                      
Independent of
  temperature, the difference between the Fermi level and
chemical potential is given by
\begin{equation}\label{e:mu}
e-\mu=E(0)(1-2n)
\end{equation}                          
 The pairing gap $\Delta$
   has a  maximum in the middle of the shell ($n=1/2$).
It is given by
\begin{equation}
\Delta=\sqrt{E^2-(e-\mu)^2}.
\end{equation}                              
Its zero temperature
value is
\begin{equation}\label{e:de0}
\Delta(0)=2E(0)\sqrt{n(1-n)}.                 
\end{equation}
 As displayed in Fig. 2, a finite temperature  $T$ decreases $\Delta$
  but its behavior as a function of $n$ is similar to that at $T=0$.
 Our calculations of $\Delta$ with all single particle levels
 inside the Debye interval give results
 practically indistinguishable from those in Fig. 2. This demonstrates
 the applicability of the single shell approximation for the
considered values of $N$ and $T$.

  The transition temperature $T_c$ is defined as the temperature where
$\Delta$ disappears. It is obtained from Eq.(\ref{e:gap}), substituting
the value of the expression (\ref{e:qp}) at $\Delta=0$ for $E$, i. e.
\begin{equation}
\mid e -\mu\mid =E(0)
\frac{\exp \{(\mid e -\mu\mid /T_c)\}-1}
{\exp \{(\mid e -\mu\mid /T_c)\}+1}.
\end{equation}                                   
Taking into account the expression (\ref{e:mu}) for the chemical potential
the transition temperature is given by
\begin{eqnarray}\label{e:tc}
T_c(n)=2E(0) \xi^{-1}(n)~,  \\
\xi (n)=2x^{-1}\ln{\frac{1+x}{1-x}}, \; \; \
x=\mid  1-2n\mid  \;,\;\  0<x<1.     \nonumber
\end{eqnarray}                                       
 As shown in Fig.3,  $T_c$ is symmetric with respect to exchanging
 $n\rightarrow 1-n$.
 In the middle of the shell ($n=1/2$, $x=0$) the
 limit $x\rightarrow 0$ in Eq.(\ref{e:tc}) gives the known result
 \cite{20}
 $\xi (n=1/2)=4$, which coincides with the value obtained
 Parameter \cite{11} for a cubic cavity.

In the  $BCS$ theory of bulk metals there is the universal
 relation between $T_c$ and $2\Delta(0)$ which is the energy of the
 lowest two-quasiparticle excitation,
\begin{equation}\label{e:tcbcs}
\frac{2\Delta (0)}{T_c}=\frac{2E(0)}{T_c}=3.52.
\end{equation}                                     
 In the single shell model at $T$=0 the two-quasiparticle
 energy $2E(0)$ is different from $2\Delta(0)$. Accordingly,
 two different ratios can be considered,
\begin{equation}\label{e:xi}                        
\frac{2E(0)}{T_c}=\frac{GM}{T_c}=\xi (n)
\end{equation}
and
\begin{displaymath}
2\Delta(0)/T_c=2\xi (n)\sqrt{n(1-n)}.
\end{displaymath}
Both depend on the shell filling parameter $n$.

\section{Averaging over the shell structure}\label{average}

The physical quantities we are interested in show rapid variations as functions
of the electron number $N$, which are manifestations of the pronounced
shell structure. 
 We will  average   some quantities $f(n)$ over
 an interval of particle numbers $N$,
\begin{displaymath}
(f(y))_{av}=\frac {1}{\gamma\sqrt\pi}
\int f(x)\exp \{ -\frac {(x-y)^2}{\gamma^2}\}dx,
\end{displaymath}
 where $x=N^{1/3}$ and $\gamma=0.3$.
The reason for averaging consists in following. Firstly,
the number of electrons is not exactly known and there is a 
considerable uncertainty in the grain radius.
 In measurement on probes
containing many grains,     
there will be an experimental $N$-distribution.
Secondly, the phases of the fast $N$-oscillations are sensitive
to small deviations from
 the ideal spherical symmetry of the
 adopted model. In real grains this
 symmetry is certainly broken by the roughness of the surface due to
the discrete ionic background, by deviations of the shape 
from a sphere or hemisphere and by impurities. 
These imperfections will shift and wash out the oscillations.
Averaging out  the shell structure oscillations has also been advocated in ref. \cite{10} 
studying the superconductivity of  electrons in a cubic cavity as well as
in the studies of the  enhancement of paramagnetism in mesoscopic 
systems (cf. e. g. \cite{richter}).   

     The
 averaged values of $2E(0)/T_c$ are shown in Fig.4. They are larger than $4$
because $T_c$   is less than $T_c(n=1/2)$ for $n\ne 1/2$
 but $E(0)$ is $n$ - independent.
 The averaged ratios of $2\Delta(0)/T_c$
are close to the bulk value 3.52 of a weakly coupled superconductor. Since
the same ratio is also found for the cubic cavity \cite{10} it may be of general
nature.

\section{ Effective pairing strength in the
$BCS$ approximation}\label{g}

The two quasiparticle energy $2E(0)$ has been measured in refs.
\cite{6,7} for different $N$ at sufficiently low temperature, such  that the
zero temperature expressions can be applied.
 Within our single shell model, eq. (\ref{e:e0} ) relates them directly
to the coupling constant $G$. In order to reproduce the experimental $N$ -
dependence of $2E(0)$ we must assume that $G(N)$ deviates from being
$\propto 1/N$ as  in the bulk. The restricted number
of data points does not allow a quite definite determination of this function.
 We have adjusted  two different
phenomenological expressions,
\begin{equation}\label{e:G1}
G=gN^{-\alpha}, \; \
g=1.94 \;  meV, \;  \  \alpha=0.47,
\end{equation}                                     
and
\begin{eqnarray}\label{e:G2}
G=gN^{-1}\exp \left[ -\alpha N^{-\beta }\right], \\
g=3.21 \; eV, \;  \  \alpha=25, \; \  \beta=0.26  \nonumber
\end{eqnarray}                                

Since often transition temperatures are measured, we show  in figs.
5a and 6a the values of $T_c$ obtained by means of (\ref{e:xi}) and
 the expressions
(\ref{e:G1}) and (\ref{e:G1}), respectively.
For comparison, the  values of $2E(0)$ measured in  refs. \cite{6,7} are
converted into "experimental" $T_c$ values by means of eq. (\ref{e:xi}),
using the averaged values of $\xi$.

 As seen in Fig.5a, expression (\ref{e:G1}) is quite reasonable
  in the region of $N \sim 10^3 \div 10^5$. Varying $\alpha$ one can
 obtain $T_c$ as an increasing ($\alpha < 1/3$) or decreasing
 ($\alpha > 1/3$) function of $N$. In the single shell model
 the particular role of $\alpha=1/3$ is connected with the
 proportionality of $T_c$ to $GM$ i.e. to $GN^{1/3}$
 ($M$ averaged is $\sim N^{1/3}$).

The $N$ - independent bulk 
coupling constant is given by $\lambda=G\rho_F$, where
 $\rho_F\propto N $ is the  density of states at  the Fermi-surface.
 Eq.(\ref{e:G2}) is constructed as a product of the bulk coupling constant and
an $N$ - dependent factor that accounts for the finite size effects.
 Accordingly,
  $g$ is estimated using the coupling constant of bulk $Al$,
 $\lambda=g\rho_F N^{-1}=0.4$ and the Fermi-gas density
 $\rho_F(Al)$. Both in
 eq.(\ref{e:G2}) and eq.(\ref{e:G1}) there are two fit parameters.
Eq.(\ref{e:G2})
 gives the correct asymptotic behavior of $G$ at
 $N\rightarrow \infty$. The particular choice of the $N$-dependence
 corresponds to  decreasing $G$ at small $N$, which is 
reflected  by the  decrease of  $T_c$ and $\Delta$
at small $N$  in Fig. 6. No physical significance is attributed 
to this decrease, because other choices of the $N$-dependence of
the factor are possible.

The function $G(N)$ obtained by fitting 
the  spherical model to the data  strongly  deviates from the bulk-law
$G \propto N^{-1}$.
It also deviates from the function $G(N)$ one would
obtain within the frame work of  models that disregard the 
shell structure, like the one of \cite{22}. 
In oder to illustrate  this statement we 
 assume equidistant levels near the Fermi surface with the spacing
$d=\rho_F^{-1}$. This model is exposed e. g. in  \cite{20}.
For $N>10^4$, 
when $\Delta (0)/d \gg 1$, it is possible to replace 
the sums over the single electron
levels by integrations and the well known expression
\begin{equation}\label{e:dbulk}
E(0)=\Delta (0)=\hbar \omega_D/\sinh(\frac{d}{G(N)})\approx 2\omega_D
\exp(-\frac{d}{G(N)})
\end{equation}
for the bulk  is obtained. 
Expressing  
the pairing constant as the product
$G(N)=d\lambda f(N)$,  the factor $f(N)$ must
grow with falling $N$
in order to reproduce the experimental observed
increase of $\Delta$ with decreasing $N$. This is at variance with our fits.
In the case of eq. (\ref{e:G1}), $f(N)=0.6N^{0.53}$  and
in the case of   eq. (\ref{e:G2}), $f(N)=\exp(-25N^{-0.26})$, which both
decrease with decreasing $N$.

Hence, if the shell structure
in small grains is not taken into account, 
the effective coupling constant must be increased
as compared to the bulk. On the other hand,
 in our spherical model the strong bunching of the electron
already  increases the value $\Delta(N)$ such that the
coupling constant must be attenuated in order to account for the more modest
increase seen in experiment. However, we do not 
consider our fits as evidence for
a decrease of the coupling constant. Due to its high symmetry, 
the spherical model
over-accentuates the shell structure. 
Imperfections of
different nature will make the level bunching in real grains less pronounced.
The spherical model and the model with equidistant levels may be considered
as the two limiting cases of maximal and no shell structure, respectively.
 The real
grains lie somewhere in between. It is quite reasonable to assume
that their shell structure is weak enough such that $f(N)$ still  increases
with decreasing $N$.   

In the context of the experimentally observed  increase of $T_c$ in thin 
films several effects have been discussed that lead to an increase
of the coupling constant  as a consequence of the reduced dimensionality
\cite{2,8,10,12,13,14,15,16}. In particular, it has been pointed out 
that the surface phonon modes in films and small grains are expected to
amplify the pair interaction. 
 Our study shows that the electronic shell
structure  must be taken into account if these effects 
are quantitatively related to the experimentally observed increase
of $T_c$, because the films often have  a granular structure.       
Even for  homogeneous films the quantization of the electron motion 
perpendicular to the surface must be taken into account, because
ref. \cite{9} has demonstrated that
the pair gap  $\Delta$ is considerably larger than its bulk value
(\ref{e:dbulk}) 
when the  thickness becomes as small as 
 few times the Fermi wave length (assuming $G=\lambda\rho_F^{-1}$).

\section{Critical temperature in finite systems}\label{can}

So far we have treated the pairing in the frame of the grand canonical
ensemble.
An import aspect of the small finite systems consists in the fact that the
number of electrons in the grain is fixed and the the canonical ensemble
must
be applied. In ref. \cite{22} this question has been investigated
on the basis of the the generalized Landau - Ginzburg equations. This
approach does not take into account the discreteness of the electron levels
and their shell structure. The spherical model
in single shell approximation is simple enough
to carry through the canonical statistics  including the  shell
structure.

As shown by Kerman \cite{23},
 the pairing
Hamiltonian can be expressed as  the Casimir operator
of the quasi spin algebra
\begin{displaymath}
H_p=-GA^+A ;\ \
A^+=\sum a^+_ta^+_{\bar {t}};\ \
\left[ A,A^+\right]=M - N_{sh} \\
\end{displaymath}
The eigenvalues $E^{(N)}_s$ are characterized by the quasi spin
$Q_s=(M-s)/2$ or seniority, which equals the  number of
unpaired particles in the shell,
\begin{displaymath}
E_s^{(N)}=-G\left[ Q_s(Q_s+1)-
\left(\frac{M -N_{sh}}{2}\right)^2-
\frac{M -N_{sh}}{2}\right]. \\
\end{displaymath}

  In this section we consider only the   particles
on the Fermi level, i. e. $N\equiv N_{sh}$. Each
eigenvalue (except the ground state one, $Q_0=M/2$) is
degenerated with the multiplicity $d_s$:

\begin{displaymath}
d_s=\left(
\begin{array}{c}
2M  \\
s
\end{array} \right)
- \left(
\begin{array}{c}
2M \\
s-2
\end{array} \right)
, \; \
d_0=1
\end{displaymath}

 The single shell model gives the energy of two quasiparticle excitations
(states with $s=2$) equal to

\begin{displaymath}
E_2^{(N)}-E_0^{(N)}=G\left[ Q_0(Q_0+1)-Q_2(Q_2+1)\right]
=GM,
\end{displaymath}

i.e. it is exactly the same value (\ref{e:e0}) as in $BCS$,  where
$GM=2E(0)$. Thus, using the exact solution does not change
the way to determine of $G$ from the empirical
 two quasiparticle energies at $T\simeq 0$  \cite{6,7}. However,
data on $T_c$ in small grains  require the consideration of the
temperature dependence of the superconductivity in
this model, which will be discussed below.

All thermodynamical quantities can be 
calculated from the canonical partition function $Z$, which is \cite{24}
\begin{displaymath}
Z_N=\sum_{s=s_ 0}^N d_s \exp{\{-(E_s^{(N)}-E_0^{(N)})/T\}},
\end{displaymath}
where
  $s_0=0$ ($s_0=1$) and $s$ are even (odd) integers if $N$ is even
(odd).

 The absence of a sharp transition from superconducting
to normal phase in finite systems can be demonstrated, for example,
by studying the temperature dependence of the internal energy $<H>$:

\begin{equation}\label{e:hp1}
<H_p>=\frac{1}{Z}
\sum d_sE_s^{(N)} \exp{\{-(E_s^{(N)}-E_0^{(N)})/T\}}.
\end{equation}                                      
It is frequently represented in the form analogous to the $BCS$ expression:
\begin{equation}\label{e:hp2}
<H_p>=-\frac{\Delta_{can}^2}{G}-E_{ex} .
\end{equation}                                     
 At $T=0$, $\Delta_{can}=G\langle N+2\mid A^+\mid N\rangle$,
where $\mid N\rangle$ is the ground
state function. It becomes $\Delta$ in the $BCS$
approximation. $E_{ex}$ is analogous to the exchange energy in
$BCS$ ( $E_{ex} (BCS)=-G\sum \langle N_i\rangle^2$). There is an arbitrariness
in the choice of this term in (\ref{e:hp2}). In ref. \cite{25}
 it is chosen such that for
half filled shell ($N=M$) it is  equal to
$-GM/4$, which  which is the temperature independent
 $BCS$ value in the single shell model.
The choice
\begin{equation}\label{e:eex}
E_{ex}=<H_p>\mid_{T\rightarrow \infty}=
\frac{\sum d_sE_s^{(N)}}{\sum d_s}=
-G\frac{N(N-1)}{2(2M-1)}
\end{equation}                                     
 seems more appropriate to the problem, because for $T\rightarrow \infty$,
 where the pairing disappears, $<H>$ given by (\ref{e:hp1})
becomes equal to (\ref{e:eex}). The difference between (\ref{e:eex}) and
 $E_{ex}(BCS)=-GN^2/4M$ is negligible at large $M$ and $N$.
 Inserting  $E_{ex}$ given by (\ref{e:eex}
) into the expression 
 (\ref{e:hp1}) for $<H_p>$,
 one gets for $T=0$
\begin{equation}\label{e:de0can}
\Delta_{can} (T=0)=\Delta (0)\sqrt {1+\frac{2}{2M-1}},
\end{equation}                      
where  $\Delta (0)$ is the $BCS$ parameter given by (\ref{e:de0}). Thus,
 $\Delta_{can}$ is larger than $\Delta (0)$ at $T=0$ but their
 ratio is practically equal to $1$ for large $M$. The quantity
 $\Delta_{can}$ exceeds $\Delta (BCS)$ at any temperature,
 especially after $T_c$ where $\Delta (BCS)=0$. As seen in Fig. 7, it
 is smoothly decreasing  and there is  no sharp transition to
 the normal state. The larger  $M$,
 the closer $\Delta_{can}$ approaches $\Delta (BCS)$.

The specific heat $C$ of the  electrons,
\begin{displaymath}
C=\frac{1}{N} T\frac{\partial^2}{\partial T^2} \{T\ln
Z\}=\frac{1}{N} \frac{\partial}{\partial T}<H>,
\end{displaymath}
becomes in the single shell model \begin{displaymath}
C=\frac{k}{N(T)^2}\left[ \frac{1}{Z}
\sum d_s(E_s^{(N)}){}^2 \exp \{ -(E_s^{(N)} -E_0^{(N)})/T \}
-<H_p>^2\right] .
\end{displaymath}
 This equation can be represented in a $G$-independent form
 if we introduce the $BCS$ critical temperature
$T_c$  as the unit of the temperature and
 energy,
\begin{displaymath}
\varepsilon_s=E_s^{(N)}/T_c, ~~~t=T/T_c,
\end{displaymath}
  where $T_c$ is  determined by (\ref{e:tc}) for given
 $M$ and $N$. As a consequence of
 $T_c$ and  $E_s$ being proportional to $G$, one has
\begin{eqnarray}\label{e:c}
C=\frac{k}{Nt^2}
\left\{
\frac{1}{Z} \sum d_s \varepsilon_s^2
\right.  \exp (
-\frac{\varepsilon_s - \varepsilon_0}{t} ) - 
\frac{1}{Z^2} \left[
\sum d_s \varepsilon_s
\exp ( -\frac{\varepsilon_s -\varepsilon_0}{t} )
\right]^2
\left. \right\}.
\end{eqnarray}     

 In the grand canonical $BCS$ approximation
  $C$ can be represented as
 follows,
\begin{displaymath}
C=\frac{1}{N}\left[ -\frac{\partial }{\partial T}
\frac{\Delta^2}{G}\right].
\end{displaymath}
 In the case $N=M$ the $BCS$ gap equation (\ref{e:gap}) takes the form
 ($t=T/T_c$)
\begin{displaymath}
\frac {\Delta}{\Delta (0)}=
\tanh (\frac{\Delta}{\Delta  (0)}\frac{1}{t}).
\end{displaymath}
In the limits $T\rightarrow  0$ and
$T\rightarrow  T_c$ one obtains for the gap and specific heat:
\begin{displaymath}
t\ll 1:  \; \ \ \frac {\Delta}{\Delta (0)}
=(1-2\exp  \{-2/t\},
\; \
\frac{C(BCS)}{k}=\frac{8}{t^2}\exp \{ -2/t\}  ,
\end{displaymath}
\begin{displaymath}
t\simeq 1: \; \   \frac {\Delta}{\Delta (0)}
=\sqrt{3(1-t)},
\; \  \frac{C(BCS)}{k}=3.
\end{displaymath}
 It is interesting to note that the jump in the
$C(BCS)$ at $t=1$  coincides with the value of
the Gorter-Casimir two-fluid model \cite{24}.

 Fig. 8 shows   $C$ calculated by means of  (\ref{e:c}). The case 
 of the half filled shell $N=M$ is shown for several
 values of $M$.
For comparison, the  $BCS$ result is added.
 Our model reproduces qualitatively
the $T$ dependence  of $C$  observed in small grains \cite{5,17}: when $N$
 decreases, the peak in $C$ shifts to smaller $t$, it becomes
 lower and  its width increases.
 A similar qualitative behavior  of $C$ was
 also obtained in the framework of the generalized Landau - Ginzburg
 model \cite{22}, and the two-level model system \cite{23}.
The similarity of the $T$ dependence of $C$ found in the three models, which
are very different with respect to the shell structure of the electron levels,
indicates that the behavior  of the function $C(T)$ shown in Fig. 8 is a
general consequence of the small and fixed number of particles. Fig.8
indicates that at $t=1$  $(T=T_c)$ the canonical
 specific heat in finite but
 not very small ($M>10$)
 systems approximately attains the half of its maximum value. This
 observation correlates with empirical definition of $T_c$ as the
 temperature at which  a measured quantity takes $0.4$ or $0.5$
 of its maximal value. This definition was used in measurements of both
 electronic specific heats \cite{5,17}
 and electromagnetic quantities \cite{16,17}. The
 critical temperatures determined this way practically
 coincide. Thus, the consideration of the theoretical specific
 heat in the single shell model canonical approach shows that the
 critical temperature calculated in the grand canonical $BCS$
 approximation can be adopted as the transition temperature $T_c$
 for small systems. Our determination of the coupling
 constant $G$ performed in the previous section on the basis of
 measured energies of two quasiparticle excitations and critical
 temperatures practically require no corrections.

\section{Magnetic susceptibility}\label{mag}

 The  magnetic susceptibility $\chi$ of the electrons in the low field
 limit is given by the change of their thermodynamic potential
\begin{equation}
\Omega (T,B) =\langle H-\mu N+\omega_L(L_z+2S_z) +
\frac {m\omega_L^2}{2}(x^2+y^2)\rangle -TS\\ 
\end{equation}
with respect to the external magnetic field $B$ (assumed to be in direction of
the z - axis), 
\begin{equation}
\chi=- \frac {1}{V}\frac {\partial {}^2\Omega}{\partial B^2}|_{B=0}.
\label{eq:chi}
\end{equation}                                      
We have introduced the Larmor frequency
$\omega_L=\mu_B B$, where $\mu_B$ and $m$ are the 
Bohr magneton and electron mass, respectively.
Here, $H-\mu N$ is the pairing hamiltonian (\ref{eq:hpair}),
$L_z$ the z -component of the orbital angular momentum, $S_z$
the z -component of the spin, 
$V$ the volume  and $S$  the entropy of the system. The single
electron levels are Zeeman-split by the magnetic field, 
\begin{eqnarray}
\varepsilon_t (B)=\tilde\varepsilon_t + \omega_L(\Lambda_t+2\Sigma_t),\\
\tilde\varepsilon_t=\varepsilon_t ^{(0)}+\frac {m\omega_L^2}{2}
\langle t\mid x^2+y^2\mid t\rangle \label{eq:r2}.
\end{eqnarray}
   The $z$-projections of the orbital
 and spin momentum are denoted by $\Lambda_t$ and $\Sigma_t$. 
The gyromagnetic factor of the electron is set to 2 and 
\mbox{$\varepsilon_t ^{(0)}=\tilde\varepsilon_t(B=0)$}.

For spherical systems
\indent
\begin{displaymath}
\langle t\mid x^2+y^2\mid t\rangle = \frac {2}{3}
\langle t\mid r^2\mid t\rangle
\end{displaymath}
and the expression for $\Omega$ can be reduced to the following form:
\begin{eqnarray}
\Omega =2\sum_{t>0} (\tilde\varepsilon_t -\mu -e_t) +\Delta^2 /G
- 2T\sum_{t,\Lambda >0} \log \left [(1+\exp (-E_t/T))
(1+\exp (-E_{\bar t}/T))\right ], \label{eq:omB}\\
\nonumber \\ \nonumber
E_t=e_t + \omega_L(\Lambda_t+2\Sigma_t), \; \; \;
E_{\bar t}=e_t - \omega_L(\Lambda_t+2\Sigma_t),  \nonumber \\
\nonumber \\ \nonumber
e_t=\sqrt {(\tilde\varepsilon_t -\mu)^2+\Delta^2}, \; \; \;
\Delta =2G\sum_{t,\Lambda >0} u_t v_t(1-f_t-f_{\bar t}), \nonumber
\\ \nonumber
f_t=(1+\exp (E_t/T))^{-1}, 
~~~f_{\bar t}=(1+\exp (E_{\bar t}/T))^{-1}. \nonumber
\end{eqnarray}                                   
For normal grains ($G=0$) the first two terms in Eq.(\ref{eq:omB}) vanish
and $E_t$ is replaced by $\tilde\varepsilon_t -\mu $. Inserting 
(\ref{eq:r2}) and (\ref{eq:omB}) into (\ref{eq:chi}),
 one obtains the grand canonical susceptibility of a spherical grain,
\begin{displaymath}
\chi =\chi_D+\chi_P,
\end{displaymath}
as a sum of  a diamagnetic and a paramagnetic
 contribution, $\chi_D$ and $\chi_P$, respectively.

The diamagnetic contribution is given by
\begin{equation}\label{eq:chid}
\chi_D=\frac {8\mu_B^2 m}{3V\hbar^2}\sum_{t,\Lambda >0}
\langle t\mid r^2\mid t\rangle n_t.
\end{equation}                                        
The occupation numbers $n_t$ are for $G>0$
\begin{eqnarray}
n_t=\frac {1}{2}\{ 1-\frac {\varepsilon_t ^{(0)}-\mu}{e_t ^{(0)}}
(1-2f_t)\}\\
e_t ^{(0)}= \sqrt {(\tilde\varepsilon^{(0)}_t -\mu)^2+\Delta^2},~~~
f_t=(1+\exp \{e_t^{(0)}/T\})^{-1}
\end{eqnarray}                                        
and for $G=0$
\begin{equation}
n_t=f_t, \; \; f_t=(1+\exp \{(\varepsilon_t ^{(0)} -\mu )/T\}^{-1}.
\end{equation}                                        
 The matrix element
 in (\ref{eq:chid}) can be straightforwardly calculated with Bessel functions,
 which are eigenfunctions of our model,
\begin{equation}\label{e:r2}
\langle t\mid r^2\mid t\rangle =
\frac {R^2}{2}\{\frac{1}{2}+\frac{(l+1/2){}^2-1} {(kR){}^2}\}{}_t.
\end{equation}
Here,  $k$ is the wave number of state $t$.
In the single shell approximation the pairing acts only in the last shell.
Since $\langle t\mid r^2\mid t\rangle$ is constant within one shell,
$ r^2 \propto N $ and equal to its value without pairing. Hence,
 $\chi_D (G,T)$ is given by (\ref{eq:chid}) calculated for $G=0$.

\indent The paramagnetic contribution,
\begin{equation}
\chi_P=\frac{\mu_B^2}{VT}\sum_{t,\Lambda,\Sigma} f_t(1-f_t)
(\Lambda_t + 2\Sigma_t)^2 =
\frac{2\mu_B^2}{VT}\sum_{t,\Lambda} f_t(1-f_t)(\Lambda_t^2 + 1),
\end{equation}                                     
 is very sensitive to
the pairing and the shell structure.
 Again, we consider small temperatures ($T<T_c$), such that the level
 spacing near the Fermi level is much larger than $T$, and
 calculate $\chi_P$ by means of the single shell model.
 Taking into account that for the level with
the number of magnetic substates $M=2l+1$,
\begin{displaymath}
\sum \Lambda^2=\frac {l(l+1)(2l+1)}{3}=\frac {(M^2-1)M}{12},
\end{displaymath}
we obtain for $\chi_P$
\begin{equation}
\chi_P(G=0,T)=\frac {\mu_B^2}{6VT}n(1-n)
M(M^2+11), \; \;
n=N_{sh}/2M,
\end{equation}                                     
\begin{equation}\label{e:chip}
\chi_P(G>0,T)=\frac {\mu_B^2}{6VT}n(1-n)M(M^2+11)
\left[ 1- \right( \frac{\Delta (T)}{\Delta (0)}\left)^2 \right].
\end{equation}                                      
As seen, the single shell model expressions for $\chi_P$ can be
written such that the influence of the pairing is expressed
by a separate factor:
\begin{equation}
\chi_P (G>0,T)=\chi_P (G=0,T)
\left ( 1- \left(\frac{\Delta (T)}{\Delta (0)}\right)^2\right ).
\end{equation}                                      





  The temperature dependence  of $\chi$  is displayed in Fig.10,
which shows the case $N=3371$,  corresponding to the middle of the shell, 
where $\chi_P$ has its maximum.
 For $T\rightarrow  0$ the paramagnetic
susceptibility goes to zero. Hence, the grain becomes an ideal diamagnet. The 
reason is the same as for an atom. The first excited state is at the
energy $2E(0)\gg T$. i. e. only the diamagnetic part (\ref{eq:chid})
contributes. The susceptibility for the non-superconducting system 
$\chi_P (G=0,T)$ diverges, because it costs no energy to occupy the
magnetic substates such that the magnetic moment is finite.   
Hence,  at small temperatures  the
susceptibility of superconducting grains is  negative. 
It increases with the temperature, changes sign and reaches its
maximum at $T=T_c$. Then it decreases again proportional to $ 1/T$.

  At low temperatures $T\sim (0.1\div 0.2)T_c$, the susceptibility 
 for the unpaired state, $\chi (G=0)$,  takes  large
  paramagnetic values in the open shells  \cite{supershell,27}.
This is an example of the general appearance of paramagnetism
in a confined electron system, which is reviewed e. g. in \cite{richter}. 
Evidence for this paramagnetic enhancement in mesoscopic normal
systems has been found \cite{levy}.    
 In our case $\chi (G=0)$ is  about $10$ times higher
 than  $\chi (G>0)$. Thus, measuring $\chi$ at these
 temperatures can give information whether a grain is
 normal or superconducting.

Averaged values of $\chi $ as a function of $N^{1/3}$
  are shown in Fig.11. Both cases $G=0$ and $G>0$ are displayed for $T=3$.
  The averaged values of $\chi (G=0)$ fluctuate around some constant.
 At $T=3 K$, which is higher than $T_c(bulk)$, many of grains
 in the range $10^3 <N< 10^5$ are superconductors. Hence, the  averaged
 values of $\chi (G>0)$ are less than $\chi (G=0)$.
 For 
 $N>10^5$ the susceptibility $\chi (G>0)$ approaches  $\chi (G=0)$.
The finite size effects, which make the small grains superconducting at 
$T=3 K$, are no longer strong enough to sustain the superconducting state. 
 For  $N^{1/3}<15$, the  two different $N$ dependences of $G$ 
are reflected by the susceptibility: 
if $G$ grows  with
   decreasing $N$ according to (\ref{e:G1}),  
$\chi$ decreases (curve $2$). If it 
 decreases according to (\ref{e:G2}), $\chi$ increases and reaches 
 $\chi (G=0)$.
 Therefore, the measurement of the $N$-dependence of the
 susceptibility at low temperatures ($T<T_c$) could give
 valuable information concerning the $N$-dependence of $G$.

Let us now derive the susceptibility for the canonical ensemble.
 In the presence of a magnetic field $B$ the degeneracy of
states with a definite seniority $s$ is lifted, i. e.
\begin{equation}\label{e:zeeman}
\delta E(B,s,L,S,\Lambda,\Sigma)=\omega_L (\Lambda_s+2\Sigma_s)+
\frac {\omega_L^2 m}{2}\langle sLS\mid \frac {2}{3} r^2\mid
sLS\rangle.
\end{equation}         
As before,  $L_s$, $S_s$, $\Lambda_s$, $\Sigma_s$ are 
the orbital and spin momenta and their projections, respectively. 
The subscript $s$
indicates that these momenta correspond to a fixed seniority $s$ and
that they are chosen  to be consistent with the Pauli principle.

In the single shell approximation, $\chi_D$ is given by (\ref{eq:chid})
calculated for $G=0$. The reason is the same as for the grand
 canonical ensemble.
 The paramagnetic contribution can be written
in a $G$-independent form like the canonical heat capacity 
if we introduce the $BCS$ critical
temperature $T_c$ as the unit of the temperature and energy,
\begin{eqnarray}
\chi_P=\frac {\mu_B^2}{ZT}\sum_{s}
\exp(-\frac {\varepsilon_s -\varepsilon_0}{t})R_s, \\
R_s=\sum_{L_s,S_s,\Lambda_s,\Sigma_s}(\Lambda_s+2\Sigma_s)^2,
\label{e:rs}\\ 
\varepsilon_s=E_s^{(N)}/T_c, ~~~t=T/T_c,
\end{eqnarray}                                     
To calculate the canonical $\chi_P$ one needs the values of the 
 orbital and spin
momenta at a given seniority. This problem is solved in the Appendix.

 The results
for the shells with $l=2$ and $l=5$ 
are shown in Fig.12 in comparison with 
the grand canonical calculations. 
The relationship (\ref{e:chip}), which is exact for the 
grand canonical ensemble,  holds with  high accuracy also in the case of
the canonical ensemble if the grand canonical $\Delta$ is replaced by
the canonical $\Delta_{can}$.
It permits a transparent interpretation of the 
modification of the susceptibility
due to the conservation of the particle number.
The consequences of the superconductivity are  expressed by  the factor
$1-  (\Delta_{can} (T)/\Delta (0))^2$. The canonical approach washes out
the sharp boundary between superconducting and normal states. As 
illustrated in fig. 7, the canonical gap disappears gradually.
Correspondingly, the transition from superconducting and normal values
of the susceptibility is smoothed. The exact calculation supports
this conclusion.

\section{Conclusions}

   We have studied the dependence of the pair correlations
 on the particle number $N$ in $Al$ nanometer-scale
 grains, $10^3<N<10^5$. Using the spherically symmetric
 infinite well as a model of the  field, we have
 calculated the  gap parameter $\Delta$ as functions of $N$.
Comparing it with experimental data from tunneling experiments,
the pairing coupling constant $G(N)$ is fixed as function of the number
of electrons $N$.

In our model,  the bunching of the electron levels (shell structure)
 strongly enhances
the pair correlations in the small grains.
This enhancement is so strong that  fitted coupling constant  
$\lambda(N) =G(N)\rho_F$ {\em decreases} with decreasing $N$, in order
 to account for the more modest increase seen in experiment.
We interpret this as a consequence that our
spherical model has  too pronounced a shell structure. The bunching of
the electronic levels in realistic grains is most likely weaker due
to deviations of the shape  from the ideal sphere, surface roughness
and impurities, Thus, the enhancement of the superconductivity due to the 
shell structure is expected to be weaker and a different 
function  $\lambda(N)$ will fit the experimental data on 
$\Delta(N)$, which for the  limiting  case of no shell structure  
{\em increases} with decreasing $N$.

The enhancement of the pair correlations by the shell structure  should be
considered as a mechanism that exists in addition to the 
increase  of the effective
pair coupling constant $\lambda$ caused by  the modification of 
the phonon spectrum in small grains.   
The completely different $N$ dependence of $\lambda$ found for
the spherical and structureless models demonstrates that  
a careful estimate of the shell structure in realistic grains is 
needed in order to determine the $N$ - dependence coupling
 constant  from the data.  

The averaged magnetic susceptibility for the spherical grains 
in the normal state  turns out to be strongly paramagnetic. 
Evidence for this paramagnetic enhancement in mesoscopic normal
systems has been found \cite{levy}.    
 In our spherical model the superconductivity strongly reduces
the paramagnetism, but the susceptibility remains large and positive. 
Thus, the surprising prediction is that small superconductors may be
paramagnetic. In realistic grains with a less pronounced
shell structure  the pair correlations may be strong enough to 
make the susceptibility negative. 
Our study shows that the susceptibility of grains composed of superconducting
material results from an intricate interplay between the pair correlations
and the spatial confinement of the electrons.

The properties of the superconducting grains with the number of atoms
below $10^5$ are significantly modified by the fixed number
of electrons. Instead of a sharp phase transition an extended
 transition region in temperature appears. A transition temperature can be  
defined as the value where the 
 the specific heat shows the most rapid drop. This
 critical temperature agrees rather  well
with $T_c$ calculated in the grand canonical $BCS$
 approximation and  can be adopted as the transition temperature $T_c$
 for small systems.

 Part of this work was performed during the visit
 of two authors (N.K. and V.M.) to
Forschungszentrum Rossendorf. They thank Prof. Prade
 for hospitality. This work is supported by the INTAS (grant
 INTAS-93-151-EXT).

\newpage
\section{   Appendix}

 The sum $R_s$ appearing in Eq.(\ref{e:rs}) amounts to the difference
of the sums of squares of orbital and spin momentum projections
corresponding to states with $N=s$ and $N=s-2$ particles,
\begin{eqnarray}
R_s=R_{N=s} -R_{N=s-2} \\
\nonumber \\
R_N=\sum_{i}\langle\psi_i^N\mid(\hat L_z+2\hat S_z)^2 \mid\psi_i^N \rangle
\nonumber
\end{eqnarray}             
$R_N$ includes the expectation values of the $z$-projection orbital
and spin momentum operators ($\hat L_z$ and $\hat S_z$). The total
number of states $\psi_i^N $ at a given $N$ is
\begin{displaymath}
\left (
\begin{array}{c}
2M  \\
N
\end{array} \right ),  \; \; M=2l+1,
\end{displaymath}
with $l$ being the orbital momentum. In what follows we
assume that $N$ is even and equal to $2k$. As known \cite{28}, the exclusion
principle requires that orbital momentum and spin functions entering
into $\psi_i$ are basic vectors of conjugate or dual representations
of the permutation group and their Young diagrams should correspond
to each other by exchange of rows and columns. The diagrams of spin
functions consist of $k+S$ squares in the first row and $k-S$ in the
second one. $S$ is the spin taking the only value for each diagram and
ranging from $0$ to $k=N/2$ for a fixed even integer $N$. The orbital
permutation symmetry is characterized by two column diagrams, the
whose length of columns  is  equal to $k+S$ and $k-S$. In general,
such diagrams involve several states with orbital momenta $L_{kS}$.
We assume that $\psi_i^N$  is a direct product of an orbital function
$\varphi(k,S,L_{kS},\Lambda)$ and a spin one $\chi(S,\Sigma)$ where
$\Lambda$ and $\Sigma$ are eigenvalues of the operators
$\hat L_z$ and $\hat S_z$ respectively.
\begin{equation}
R_{N=2k}=\sum_{L_{kS},S,\Lambda,\Sigma}
\langle\varphi (k,s,L_{kS})\chi(S,\Sigma)\mid
(\hat L_z+2\hat S_z)^2\mid\varphi (k,s,L_{kS})\chi(S,\Sigma)\rangle
\end{equation}                           
As the sums over $\Lambda$ and $\Sigma$ are independent,
$R_N$ can be divided into orbital and spin parts:
\begin{equation}
R_N=R_N^{\Lambda} +R_N^{\Sigma},
\end{equation}                           
\begin{equation}
R_{N=2k}^{\Lambda}=\sum_{L_{kS},S}
\langle\varphi (k,s,L_{kS})\mid
\hat L_z^2\mid\varphi (k,s,L_{kS})\rangle (2S+1),
\label{e:a4}
\end{equation}                           
\begin{equation}\label{e:a5}
R_{N=2k}^{\Sigma}=\sum_{S,\Sigma}
\langle\chi(S,\Sigma)\mid
4\hat S_z^2\mid\chi(S,\Sigma)\rangle d_{kS}=
\frac {4}{3}\sum_{S=1}^{k} S(S+1)(2S+1)d_{kS},
\end{equation}                           
where
$d_{kS}$ is the dimension of the orbital space with the permutation
symmetry described by the two column Young diagrams mentioned above.
\begin{equation}
d_{kS}=\left (
\begin{array}{c}
M+1  \\
k-S
\end{array} \right )
\left (
\begin{array}{c}
M+1 \\
k+S
\end{array} \right  )
\frac {2S+1}{2M+1}
\end{equation}                           
\begin{equation}
\sum_{S=0}^{k}(2S+1)d_{kS}=
\left (
\begin{array}{c}
2M  \\
2k
\end{array} \right )
\end{equation}                           
Eq.(\ref{e:a5}) indicates that $R_{N}^{\Sigma}$ can be computed
straightforwardly whereas finding $R_{N}^{\Lambda}$ needs
the determination of the set of $L_{kS}$.\\
\indent This task can be removed by taking into account that external
product of two completely antisymmetric functions
$\tilde\varphi (k+S,L,\Lambda)$ ($k+S$ squares in the only column of the
Young diagram) and $\tilde\varphi (k-S,L',\Lambda ')$ ($k-S$ squares)
gives rise to a series of basic vectors of irreducible
representations with two column diagrams. Each vector arises only once
and the lengths of columns vary from $(k+S,k-S)$ up to ($2k,0$), i.e.
the decomposition of this external product contains
$\varphi (k,S',L_{kS},\Lambda)$ functions with $S'_{min}=S$ and
$S'_{max}=k$. This decomposition permits to reduce Eq.(\ref{e:a4}) to the
following

\begin{displaymath}
R_{N=2k}^{\Lambda}=\sum_{S=0}^{k} \sum_{L,\Lambda}
\left \{
\langle\tilde\varphi (k+S,L,\Lambda)\mid
\hat L_z^2\mid\tilde\varphi (k+S,L,\Lambda)\rangle
\right (
\begin{array}{c}
M  \\
k-S
\end{array} \left ) + \right.
\end{displaymath}

\begin{equation}
\langle\tilde\varphi (k-S,L,\Lambda)\mid
\hat L_z^2\mid\tilde\varphi (k-S,L,\Lambda)\rangle
\left (
\begin{array}{c}
M  \\
k+S
\end{array} \right )
\left \} \right. (2-\delta_{S,0})
\end{equation}
The orbital momenta $L$ of $\tilde\varphi $ have to be compatible
with antisymmetry of these states. This points out an elementary way
to compute them by summing single particle projections among which should
not be identical ones.

\newpage

\begin{figure}[t]
\mbox{\psfig{file=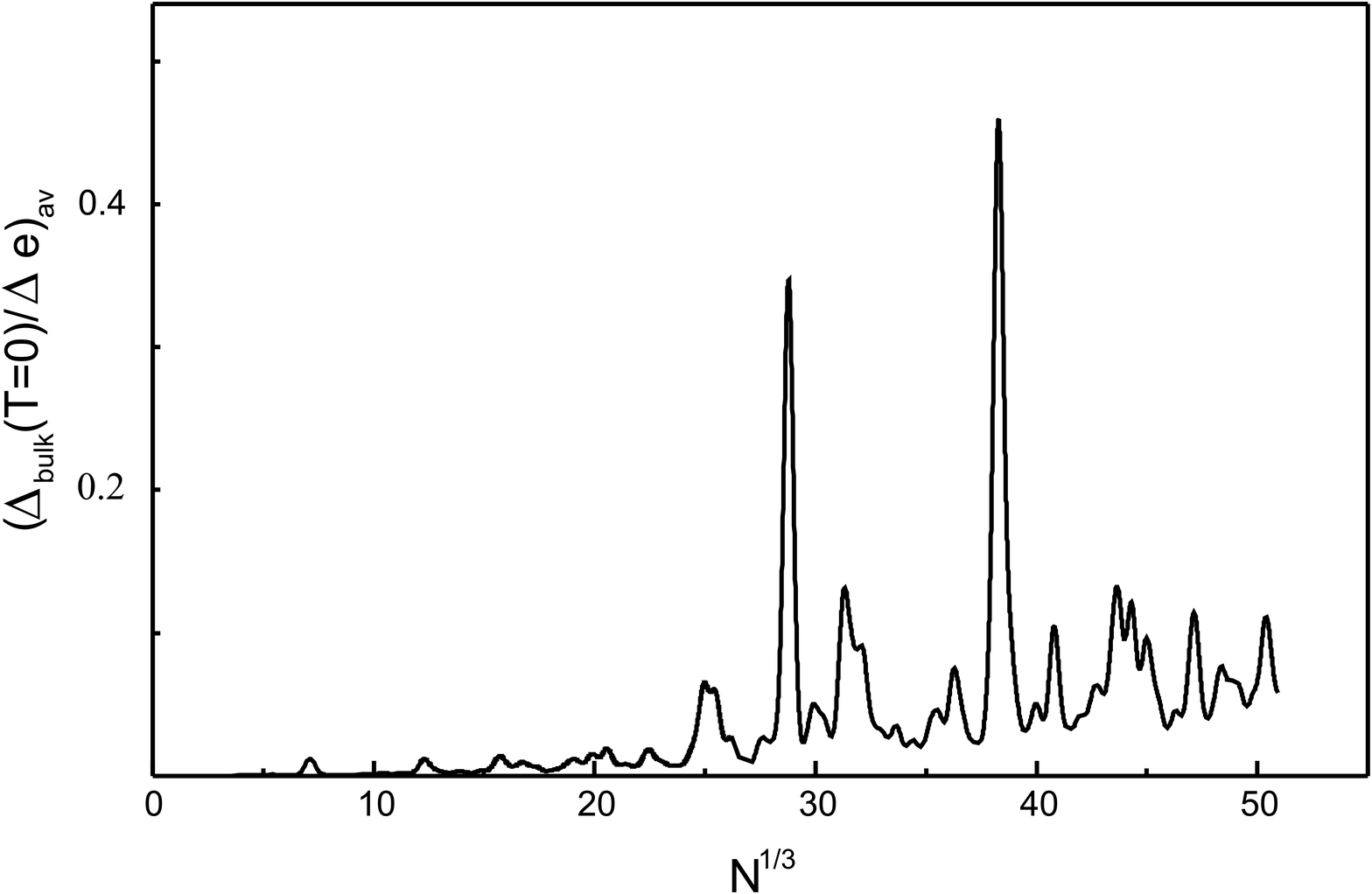,width=14cm}}
 \caption{
\label{fig1} Averaged values of $\Delta_{bulk}(T=0)/ \Delta e$
v.s. $N^{1/3}$. The pairing gap of bulk $Al$ is  $\Delta_{bulk}(T=0)$ and
 $\Delta
e=e_{F+1}-e_F$. $e_F$, $e_{F+1}$ are energies of
the Fermi level and the next higher one. The meaning of the averaging is
 explained at the end of section \protect \ref{average}. }
\end{figure}

\newpage
\begin{figure}[t]
\mbox{\psfig{file=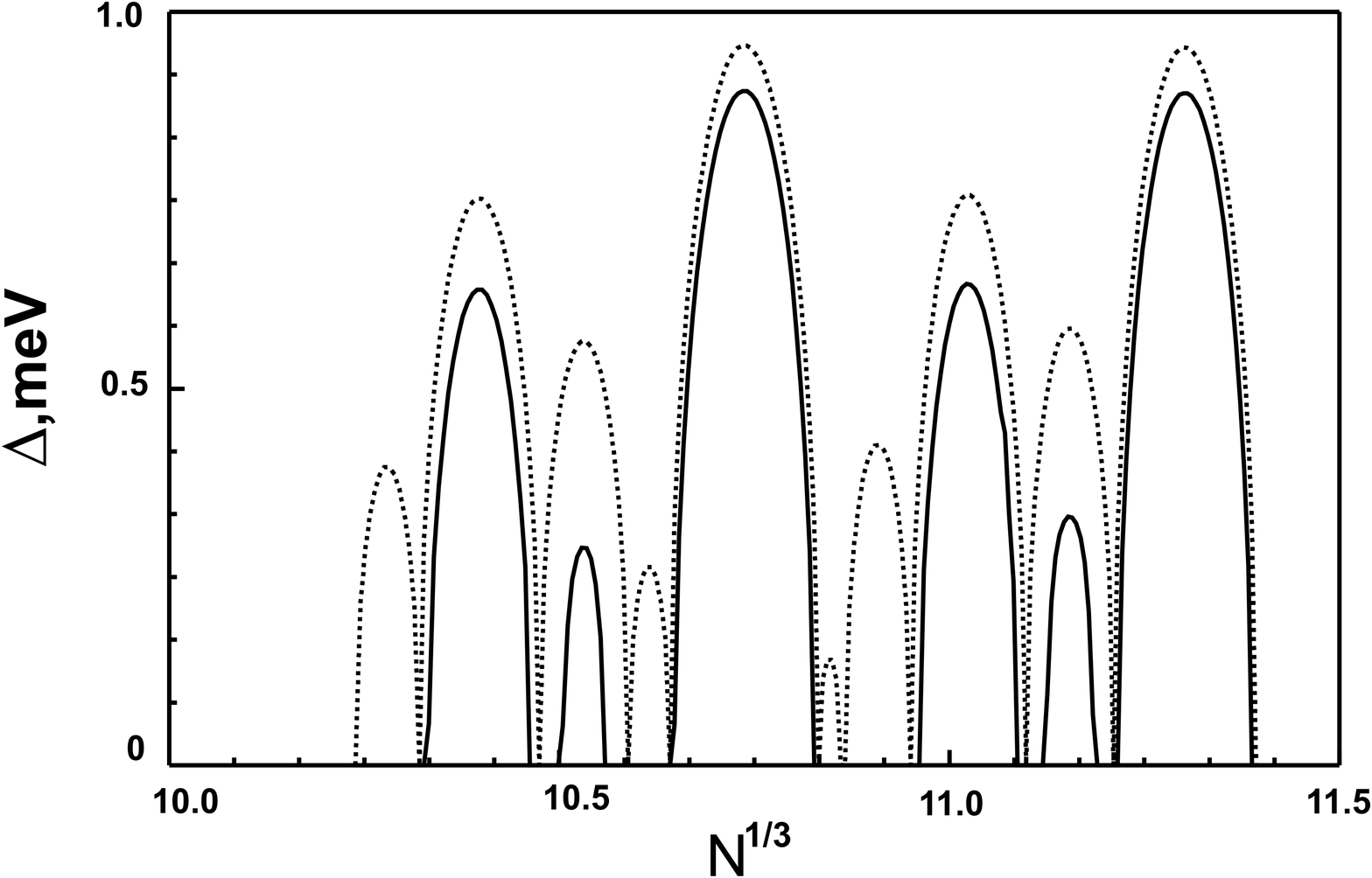,width=14cm}}

\caption{ \label{fig2} Pairing gap $\Delta (T,N)$ v.s. $N^{1/3}$.
Dashed lines: $T=0$, solid lines: $T=3$ K.}

\end{figure}
\newpage
\begin{figure}[t]
\mbox{\psfig{file=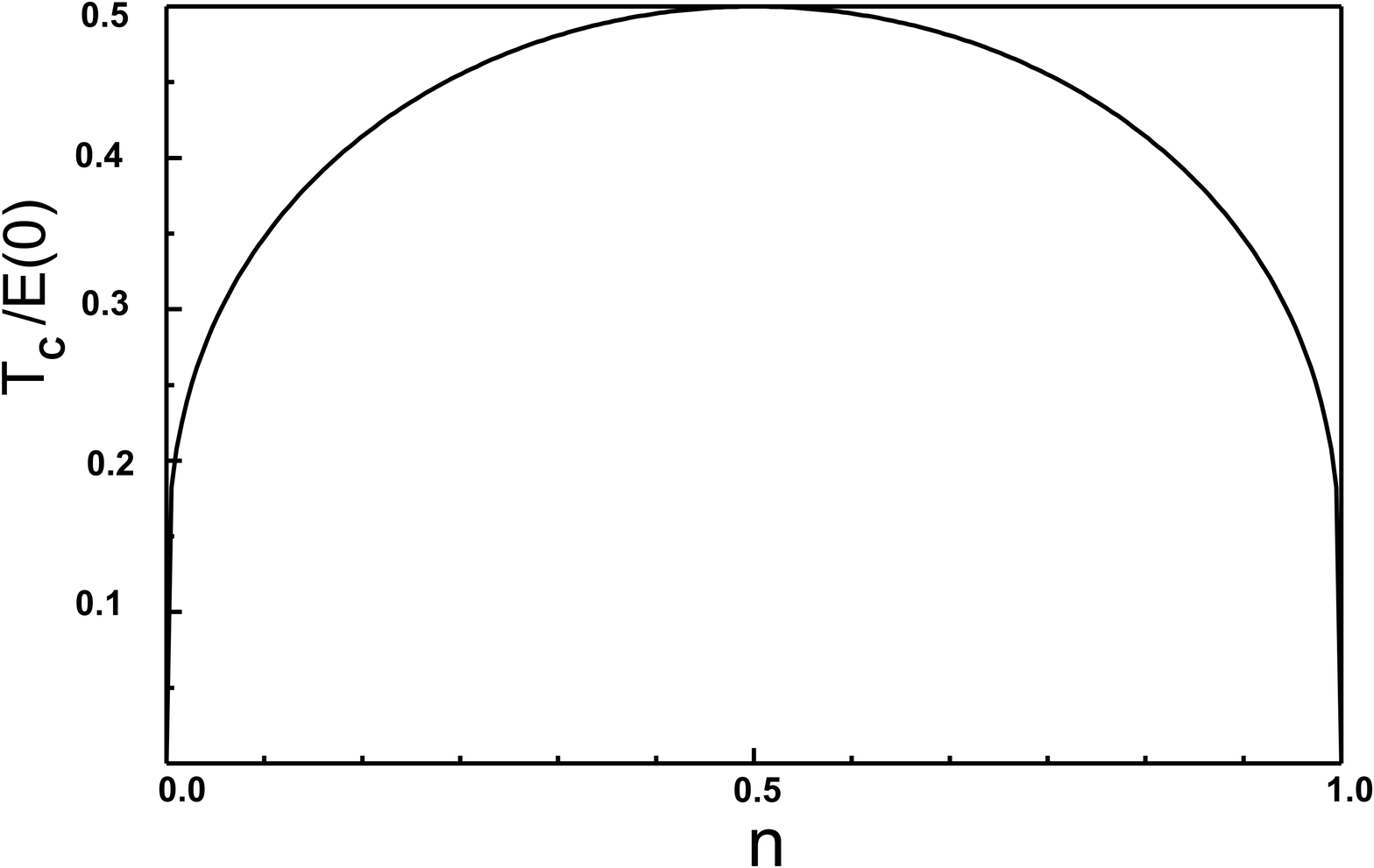,width=14cm}}
\caption{ \label{fig3} $T_c/E(0)$ v.s. the occupation degree of
the shell $n=N_{sh}/2M$.}

\end{figure}
\newpage
\begin{figure}[t]
\mbox{\psfig{file=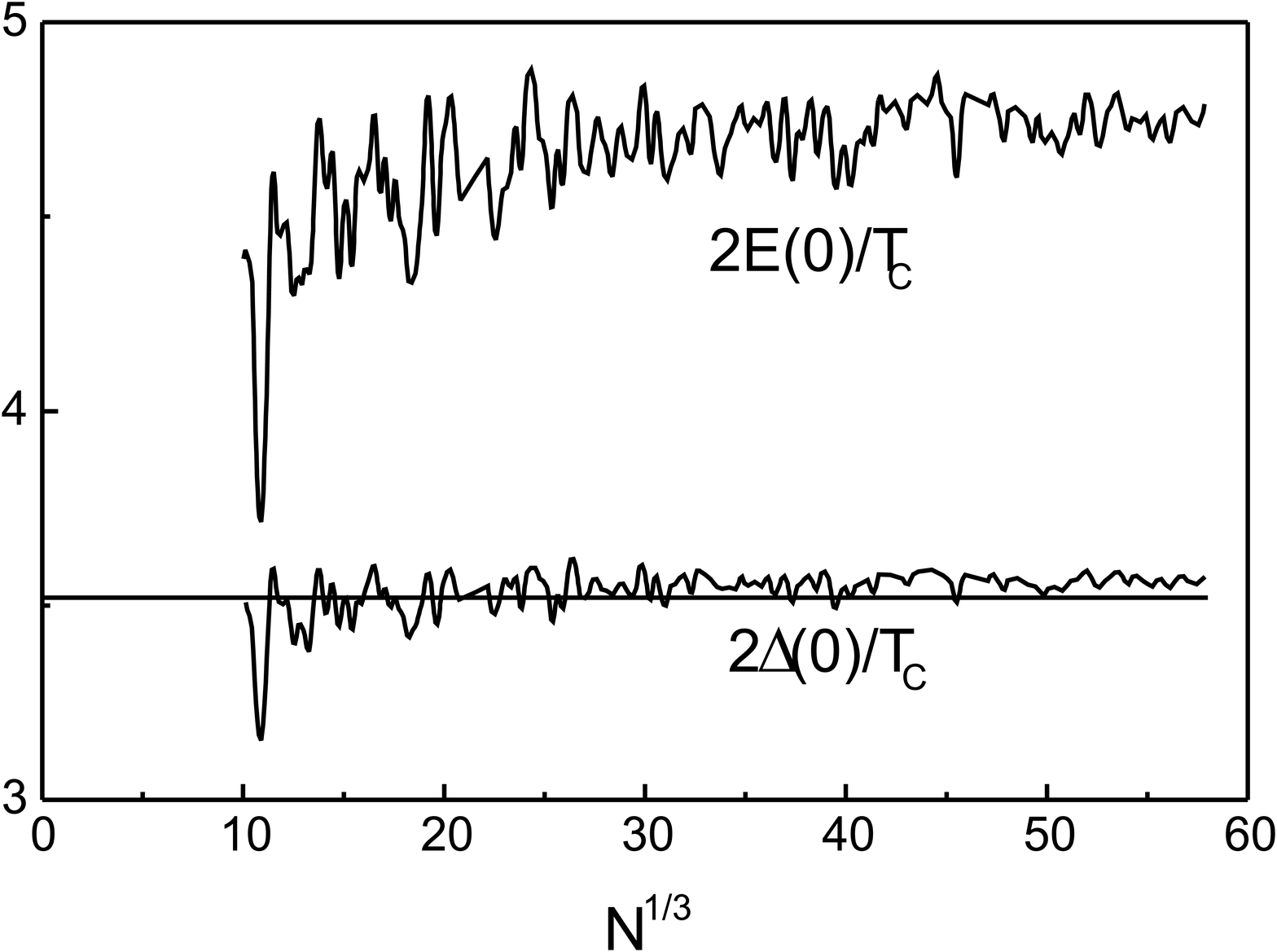,width=14cm}}
\caption{ \label{fig4} Averaged values of $2E(0)/T_c$ and
$2\Delta (0)/T_c$ v.s. $N^{1/3}$.}

\end{figure}
\newpage
\begin{figure}[t]
\mbox{\psfig{file=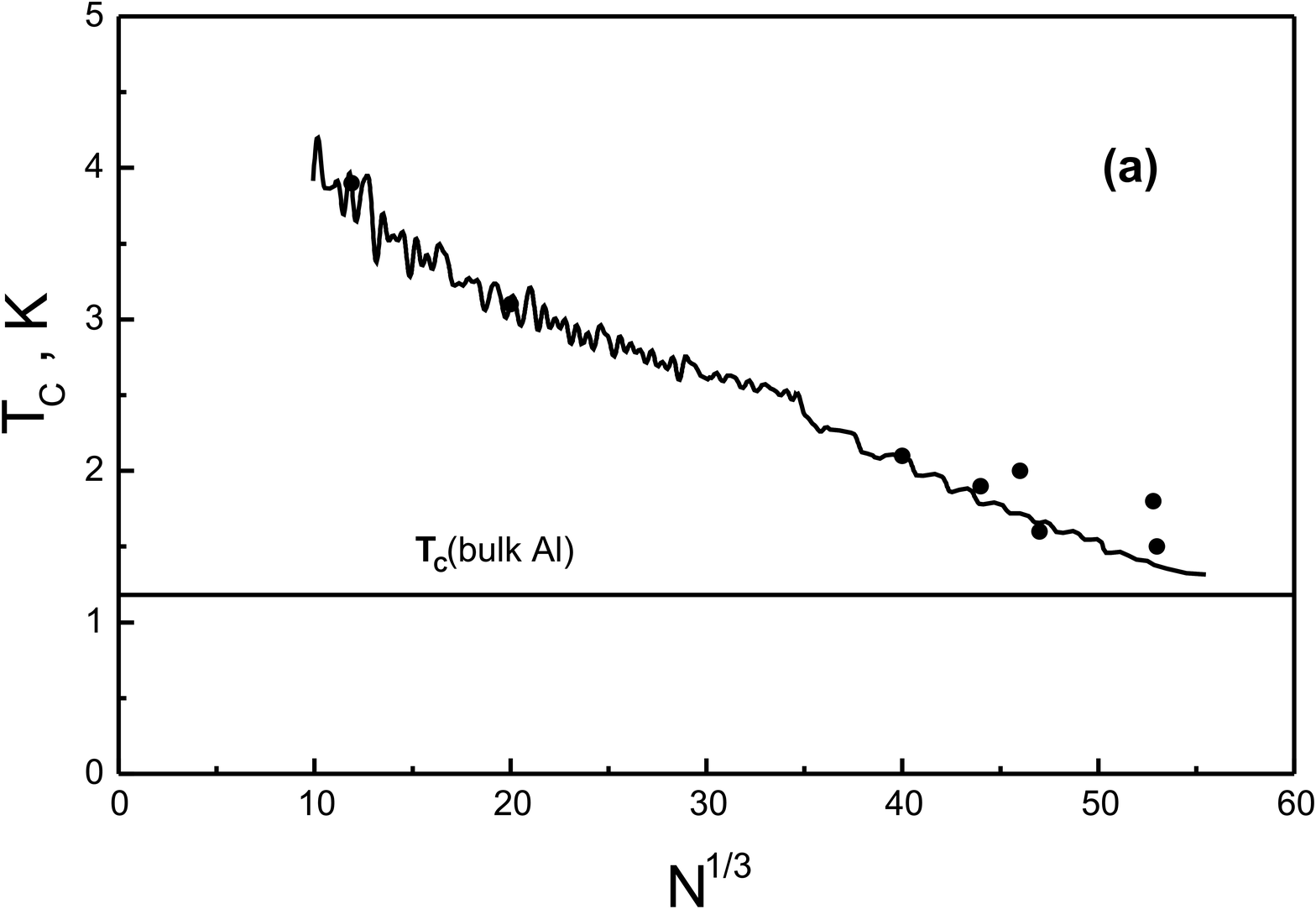,width=14cm}}
\mbox{\psfig{file=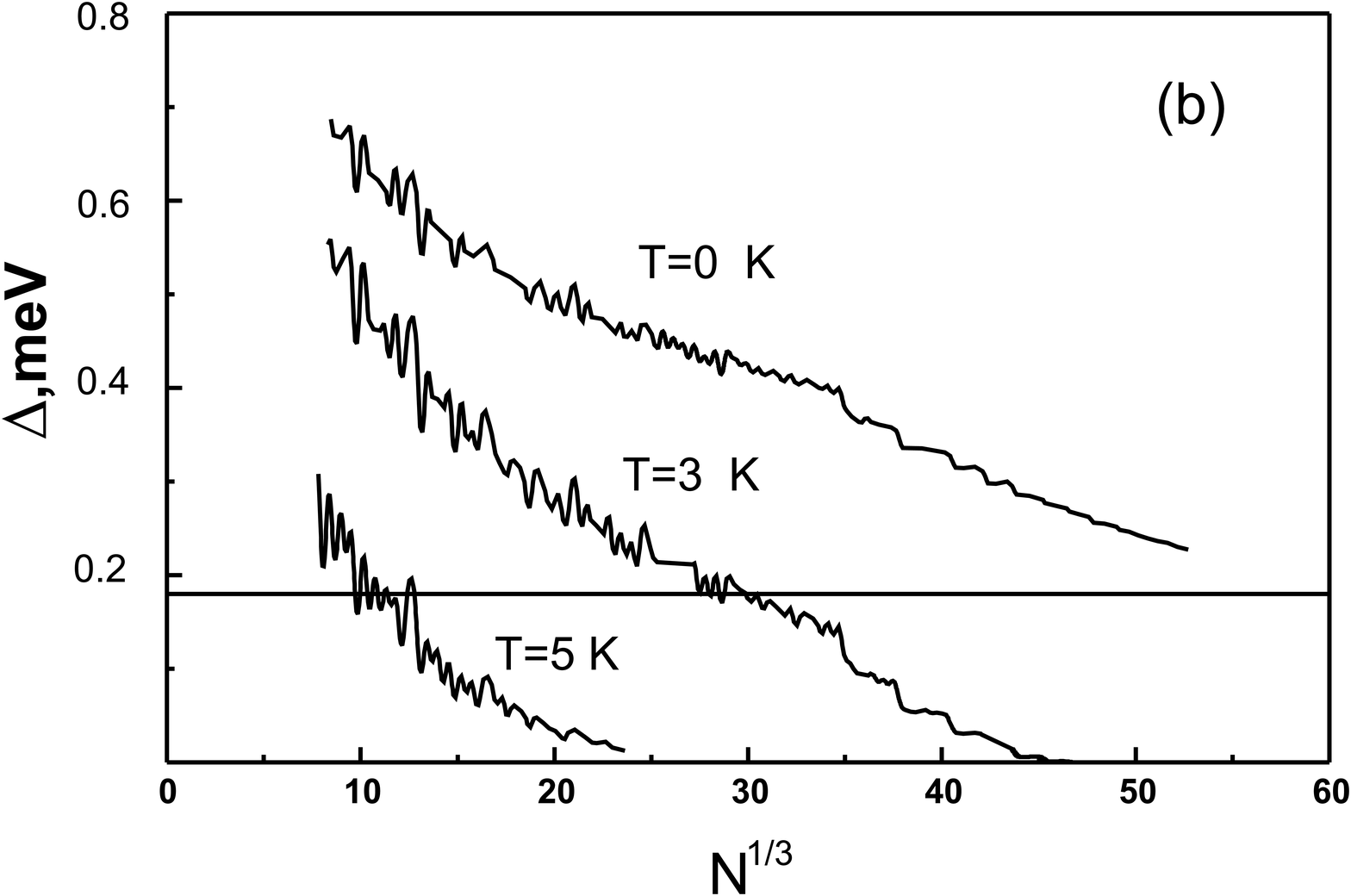,width=14cm}}
\caption{ \label{fig5} (a) $T_c$ v.s. $N^{1/3}$. Points represent
 experimental data \protect\cite{5,6} on $\Delta$, which are converted  
by mean of eq. (\protect\ref{e:xi}) into $T_c$.
The solid line is the averaged $T_c$
 calculated with $G=1.94N^{-0.47}$  $meV$.
(b) Averaged pairing gap v.s. $N^{1/3}$ calculated at $T=0,3$ $K$
 and $5$ $K$. The horizontal line gives $\Delta (0)$ of bulk $Al$.}
\end{figure}

\newpage
\begin{figure}[t]
\mbox{\psfig{file=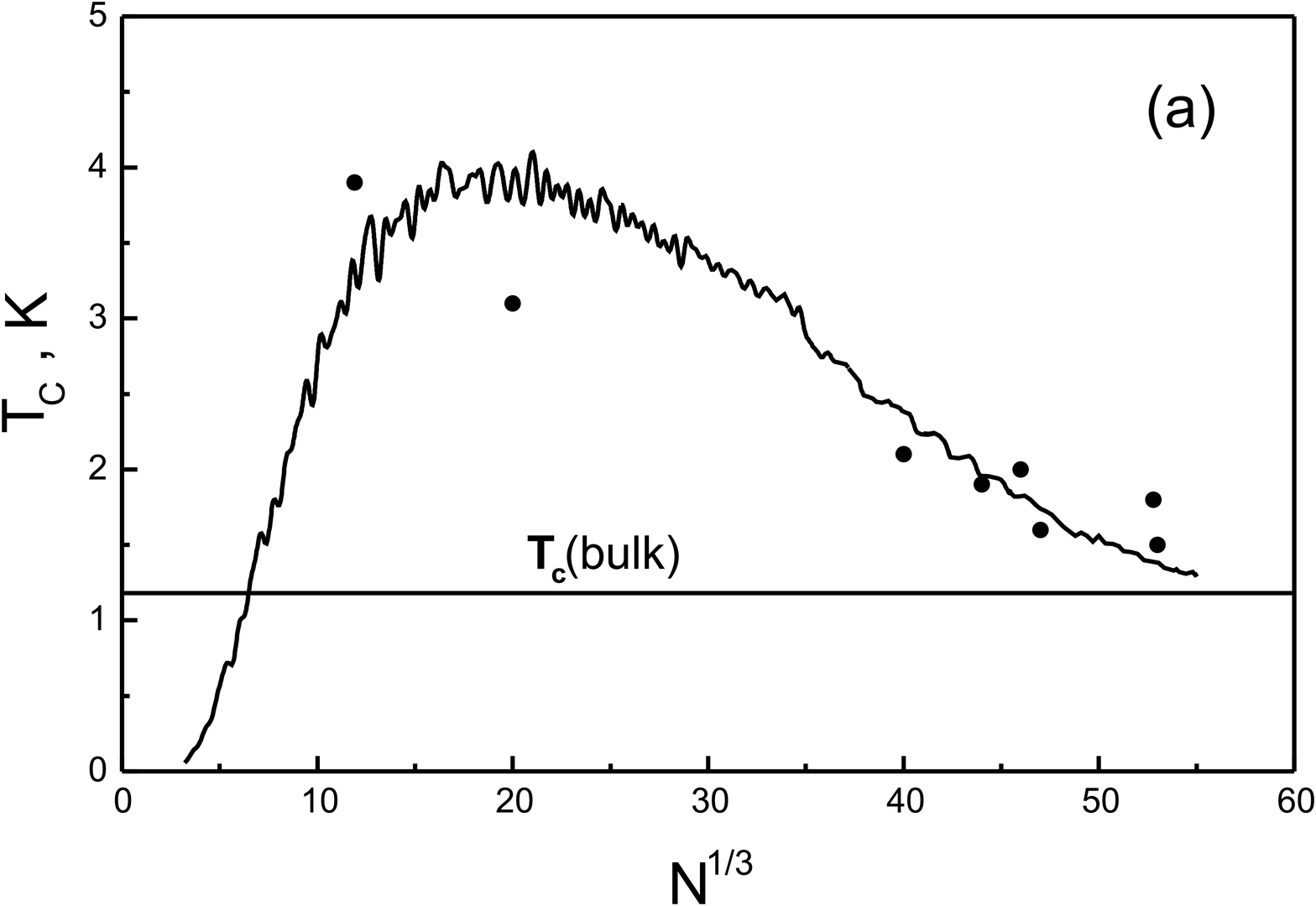,width=14cm}}
\mbox{\psfig{file=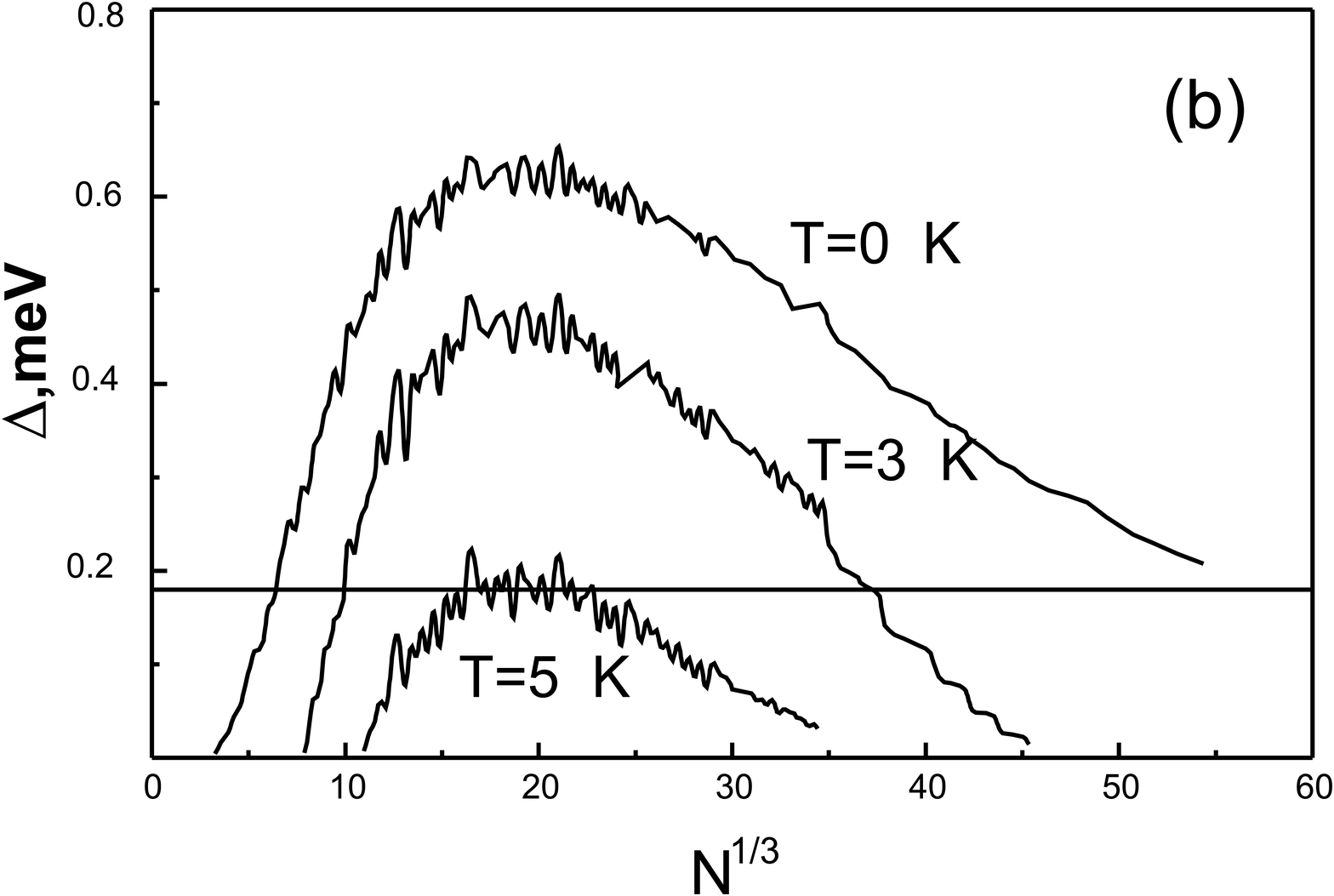,width=14cm}}
\caption{ \label{fig6} (a) $T_c$ v.s. $N^{1/3}$. Points represent
experimental data \protect\cite{5,6} on $\Delta$, which are converted  
by mean of eq. (\protect\ref{e:xi}) into $T_c$.
 The solid line is the averaged
$T_c$ calculated with $G=3.21N^{-1}\exp (-25N^{-0.26})$ $eV$.
(b) See the caption in \protect\ref{fig5}b.}

\end{figure}
\newpage
\begin{figure}[t]
\mbox{\psfig{file=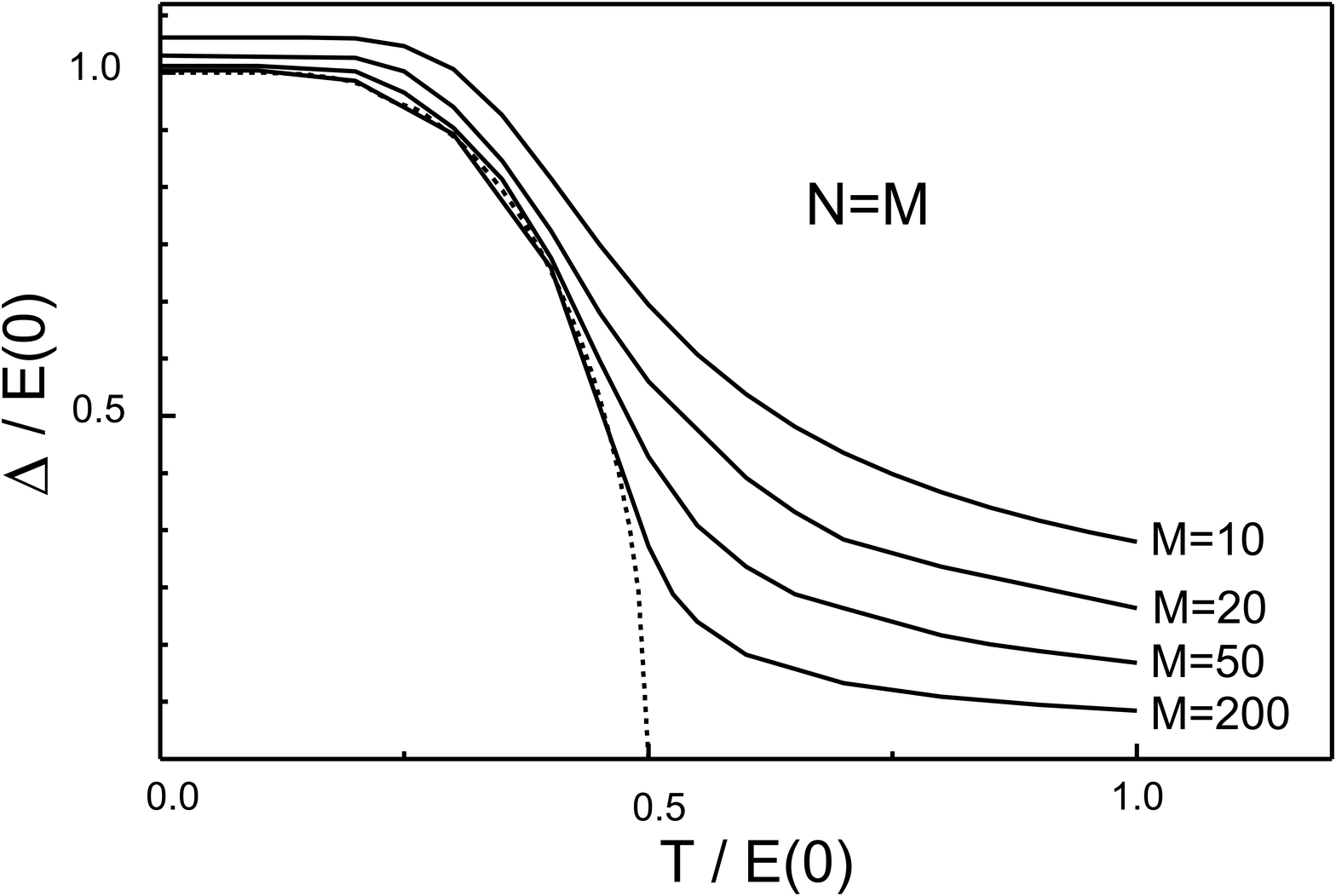,width=14cm}}
\caption{ \label{fig7} Pairing gaps calculated  v.s. $T/T_c$ at
different values of $M$ for the half filled shell.  
Solid and dashed lines correspond to
canonical and $BCS$ results,   respectively.}

\end{figure}
\newpage
\begin{figure}[t]
\mbox{\psfig{file=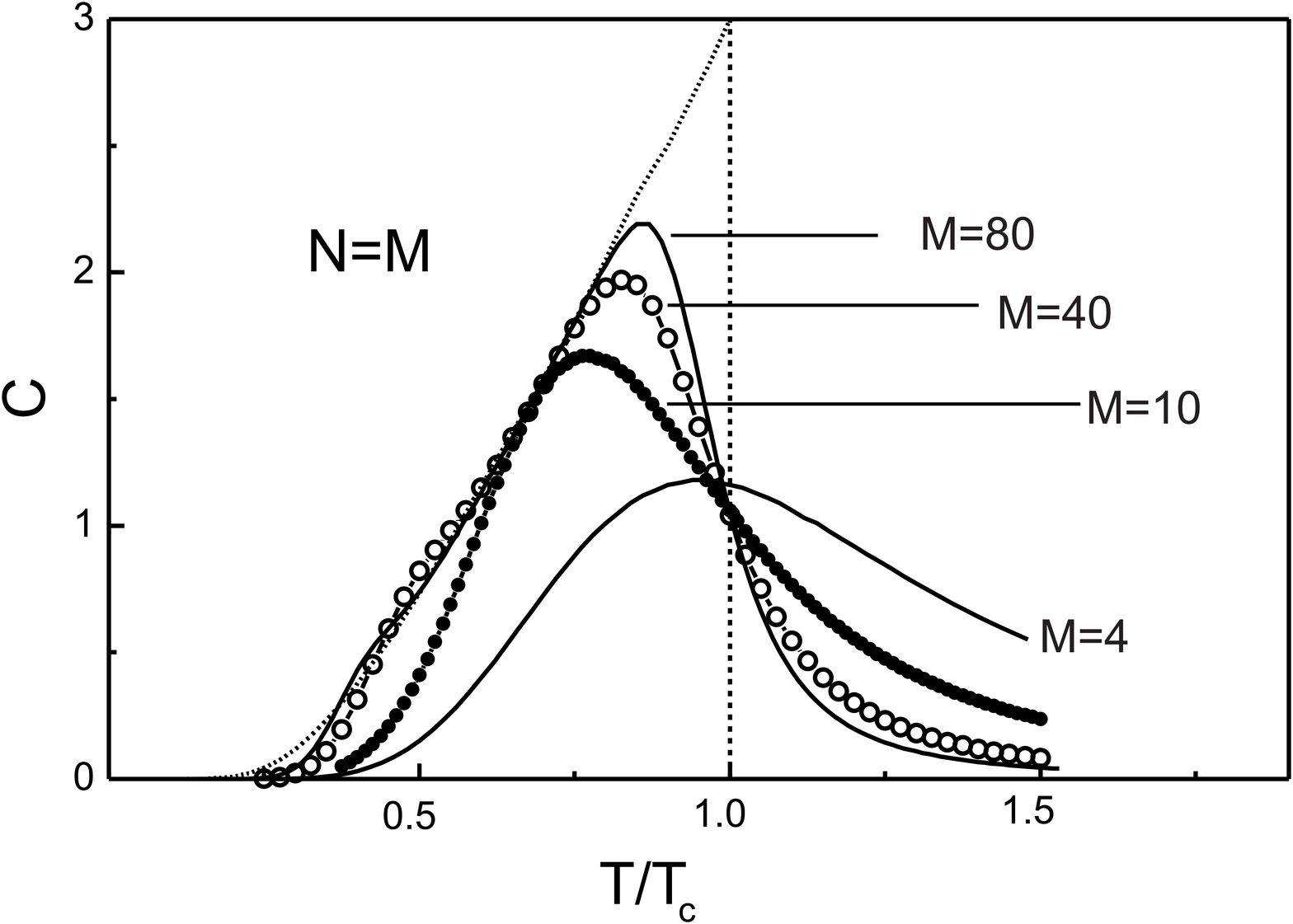,width=14cm}}
\caption{ \label{fig8} Specific heat $C$ of the single shell v.s. $T/T_C$
at
different $M$.
Solid and dashed lines correspond to
canonical and $BCS$ results, respectively.}

\end{figure}
\newpage
\begin{figure}[t]\label{fig9}
\mbox{\psfig{file=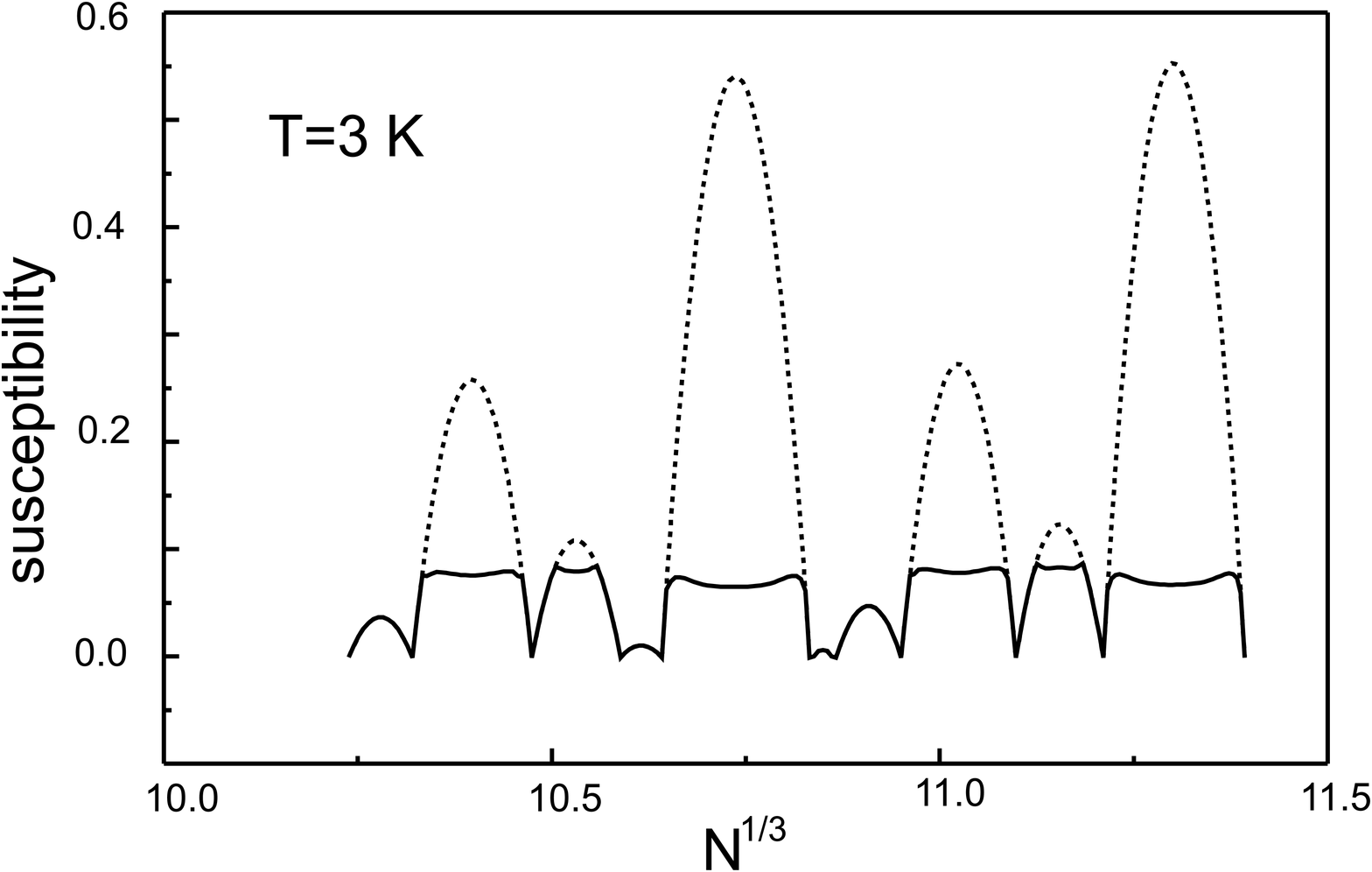,width=14cm}}
\caption{  Susceptibility
(in units $\mid \chi_L \mid \cdot 10^5$ where $\chi_L$ is the Landau
expression for diamagnetism  of the degenerate free electron gas)
v.s. $N^{1/3}$.
Dashed lines: $G=0$, solid lines:$G>0$.}

\end{figure}
\newpage

\begin{figure}[t]
\mbox{\psfig{file=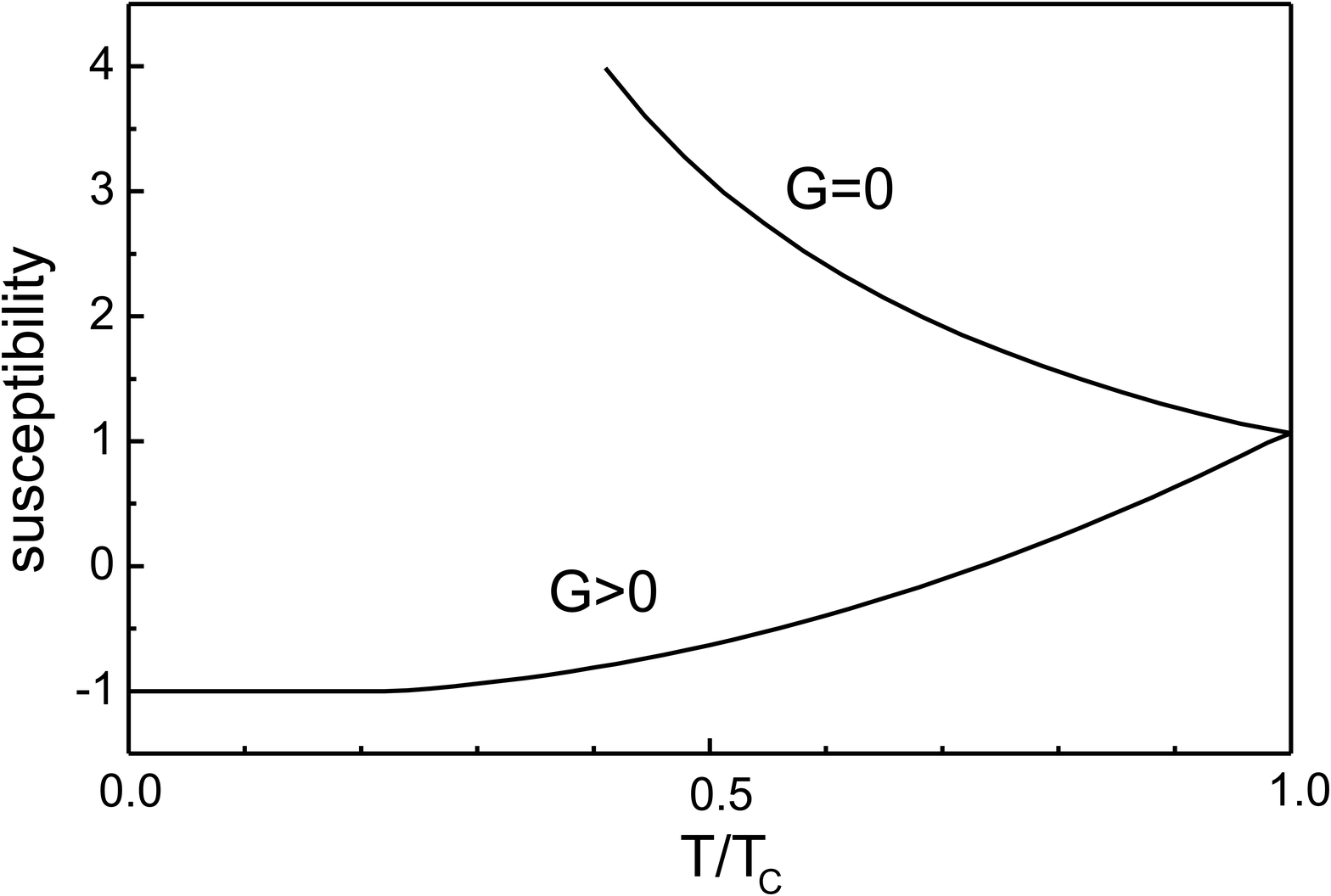,width=14cm}}
\caption{ \label{fig10} Susceptibility
(in units $\mid \chi_D \mid (G=0,T=0)$) v.s. $T/T_c$.
Dashed and solid lines give $\chi$ at $G=0$ and
$G>0$, respectively.}

\end{figure}
\newpage
\begin{figure}[t]
\mbox{\psfig{file=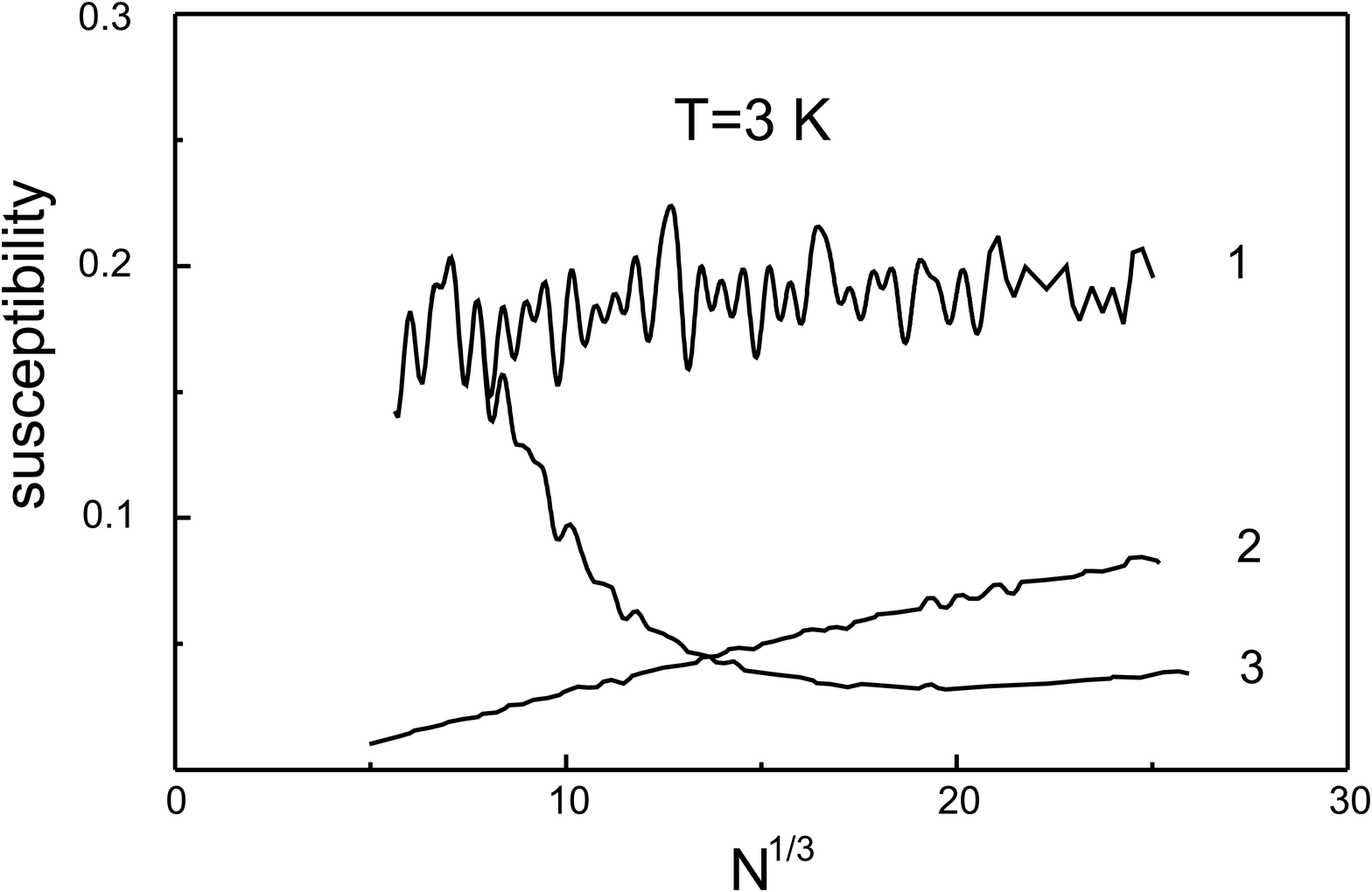,width=14cm}}
\caption{ \label{fig11} Averaged susceptibility (in units $\mid \chi_L
\mid \cdot 10^5$) v.s. $N^{1/3}$ at $T=3$ K.
Curves $1,2,3$ correspond to $G=0$;
$G=1.94N^{-0.47}$  $meV$;  $G=3.21N^{-1}\exp (-25N^{-0.26})$ $eV$,
respectively.}

\end{figure}
\newpage
\begin{figure}[t]
\mbox{\psfig{file=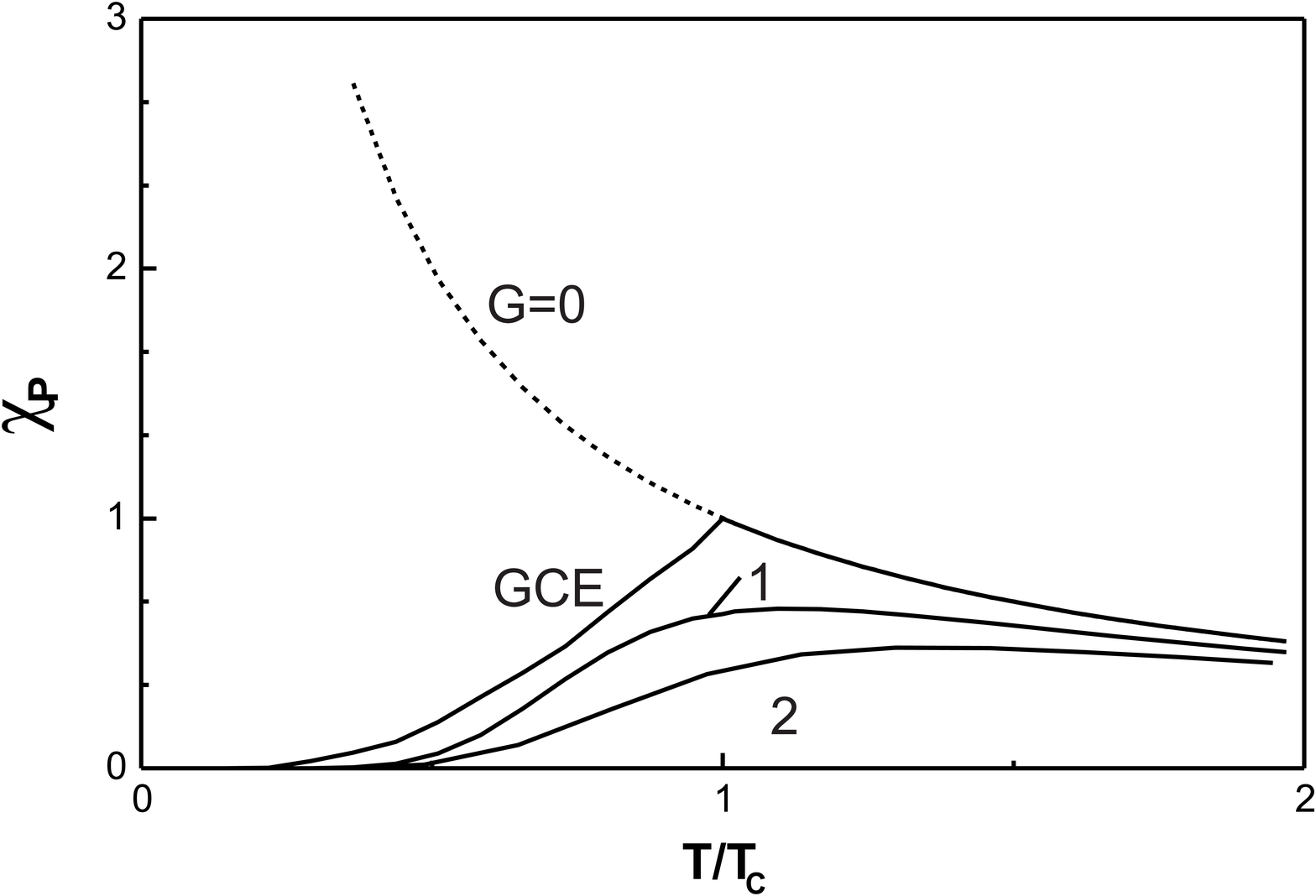,width=14cm}}
\caption{ \label{fig12} Grand canonical and canonical
 paramagnetic susceptibilities
 (in units $\chi_P (G=0,T=T_c)$ ) v.s. $T/T_c$. Curves 1,2 represent
  the results of canonical calculations of $\chi_P (G>0)$ for the
 shell with $M=11$, $N_{sh}=10$ (curve 1) and for the shell
  with $M=5$, $N_{sh}=5$ (curve 2).}
\end{figure}
\end{document}